\documentclass[useAMS,usenatbib]{mn2e}  

\usepackage{dsfont}
\usepackage{amsmath}
\usepackage{amssymb}
\usepackage{graphicx}
\usepackage{verbatim}
\usepackage{natbib}
\usepackage{eucal}
\usepackage{calligra}

\usepackage{amsfonts}
\usepackage{color}
\usepackage[normalem]{ulem}
\usepackage[T1]{fontenc}

\DeclareMathAlphabet{\mathscr}{OT1}{pzc}%
                                 {m}{it}
                                 
\newcommand{\mnras}{MNRAS}
\newcommand{\jcap}{JCAP}
\newcommand{\apj}{ApJ}
\newcommand{\prd}{Phys. Rev. D}
\newcommand{\nat}{Nature (London)}

\newcommand{\be}{\begin{equation}}
\newcommand{\ee}{\end{equation}}
\newcommand{\bes}{\begin{equation*}}
\newcommand{\ees}{\end{equation*}}
\newcommand{\bea}{\begin{eqnarray}}
\newcommand{\eea}{\end{eqnarray}}
\newcommand{\beas}{\begin{eqnarray*}}
\newcommand{\eeas}{\end{eqnarray*}}

\newcommand{\msun}{M_{\odot}}

\newcommand{\de}{{\rm d}}

\newcommand{\rhoin}{\rho_{\rm in}}
\newcommand{\rhoout}{\rho_{\rm out}}
\newcommand{\phiin}{\phi_{\rm in}}
\newcommand{\phiout}{\phi_{\rm out}}

\newcommand{\lamin}{\lambda_{\rm in}}
\newcommand{\lamout}{\lambda_{\rm out}}

\newcommand{\mpl}{M_{\rm Pl}}

\newcommand{\senv}{S_{\rm env}}
\newcommand{\renv}{R_{\rm env}}
\newcommand{\denv}{\delta_{\rm env}}
\newcommand{\dsc}{\delta_{\rm sc}}
\newcommand{\dv}{\delta_{\rm v}}

\begin{document} 

\title[Voids in Modified Gravity]
  {Voids in Modified Gravity: Excursion Set Predictions}
\author[J. Clampitt et al.]
  {Joseph~Clampitt$^{1,}$\thanks{E-Mail: clampitt@sas.upenn.edu} Yan-Chuan~Cai$^{1,2,}$\thanks{E-mail: y.c.cai@durham.ac.uk}, Baojiu~Li$^{2,}$\thanks{E-mail: baojiu.li@durham.ac.uk}\\
  $^1$Department of Physics and Astronomy, University of Pennsylvania, 209 S. 33rd St., Philadelphia, PA 19104, USA\\
  $^2$Institute for Computational Cosmology, Department of Physics, University of Durham, South Road, Durham DH1 3LE, UK}
\maketitle

\begin{abstract}
We investigate the behavior of the fifth force in voids in chameleon models using the spherical collapse method.  
Contrary to Newtonian gravity, we find the fifth force is repulsive in voids.
The strength of the fifth force depends on the density 
inside and outside the void region as well as its radius. It can be many times larger than 
the Newtonian force and their ratio is in principle unbound. This is very different from the case in halos, 
where the fifth force is no more than 1/3 of gravity. 
The evolution of voids is governed by the Newtonian gravity, the effective dark energy force 
and the fifth force. While the first two forces are common in both $\Lambda$CDM and chameleon universes,
the fifth force is unique to the latter. Driven by the outward-pointing fifth force, individual voids in chameleon models expand faster and grow 
larger than in a ${\rm \Lambda}$CDM universe. The expansion velocity of the void shell can be $20\%$ to 
$30\%$ larger for voids of a few Mpc$/h$ in radius, while their sizes can be larger by $\sim 10\%$. This 
difference is smaller for larger voids of the same density. We compare void statistics using excursion set theory; 
for voids of the same size, their number density is found to be larger in chameleon models.  
The fractional difference increases with void size due to the steepening of the void distribution function. 
The chance of having voids of radius $\sim 25$ Mpc/$h$ can be 2.5 times larger. This difference is about 10 times 
larger than that in the halo mass function. We find strong environmental dependence of void properties and population 
in chameleon models. The differences in size and expansion velocity with GR are both 
larger for small voids in high density regions. In general, the difference between chameleon models 
and ${\rm \Lambda}$CDM in void properties (size, expansion velocity and distribution function) are larger 
than the corresponding quantities for halos. This suggests that voids might be better candidates than halos for testing gravity.
\end{abstract}


\section{Introduction}

Models of modified gravity (MG) are introduced to explain the observed accelerating cosmic expansion, 
without invoking a cosmological constant in the Einstein equation.
Scalar-tensor
gravity theories are among those that are well received recently. In
these theories, the scalar field is coupled to matter,
triggering an extra fifth force which leads to an universal
enhancement of gravity. The enhanced gravity violates existing robust tests of general relativity (GR) in the solar system, so that only theories with a $screening$ $mechanism$
to suppress the fifth force in high density
regions are observationally viable \citep[e.g.][]{kw}. Gravity is therefore back to GR in
the early universe, as well as in the vicinity of virilized
objects where the local density is sufficiently high. MG models like
chameleon gravity can therefore pass the tests of current constraints from the solar 
system \citep{kw}. Nevertheless, structure formation in these models should
be somewhat different from that of the standard $\Lambda$-cold-dark-matter ($\Lambda$CDM, where $\Lambda$ represents the cosmological constant) paradigm. In low
density regions of the universe, the fifth force is weakly or not suppressed, so that
dark matter and ordinary matter will feel this extra force and hence
evolve differently from the GR case. Qualitatively, one may expect
structure to form earlier in MG than in GR with the help of
enhanced gravity. Indeed, halos are found to be more
massive and more abundant in simulations of $f(R)$ gravity
\citep{lzk2012} as compared to GR at the same epoch. Similarly,
voids appear to be larger and emptier in MG.
These qualitative results seem to point in the same direction
as some recent observational facts, which have been shown to be in tension
with a $\Lambda$CDM universe.

Firstly, some galaxy clusters detected using X-ray and lensing
techniques at high redshift are found to be too massive and
have formed too early \citep[e.g.][]{Enqvist11,Jee11,Hoyle2011,Holz2012}. 
The probability of the existence of
those massive clusters in $\Lambda$CDM is prohibitively small, but see 
\citet{Hotchkiss2011,Harrison2012, Hoyle12,Waizmann2012a, Waizmann2012b}.
Introducing non-Gaussianity can ease this tension, but the $f_{\rm NL}$ parameter required to fit the
data is usually too high, which is in tension
with other observational constrains like the cosmic microwave background (CMB).
Secondly, the detected integrated Sachs Wolfe (ISW) \citep{Sachs1967} signal from the stacking of
4-deg$^2$-size regions of the CMB corresponding to the SDSS super
clusters and super voids is found to be $2-3 \sigma$ higher
than estimations from simulations \citep{Granett08, Papai11}. This
tension with the $\Lambda$CDM paradigm is perhaps more
than $3-\sigma$ as suggested in \citet{Nadathur12}. 
Accounting for non-linear ISW effect by using simulations 
of full-sky ISW maps from \citep{Cai2010}, the tension remains nearly unchanged 
\citep{Flender2012}. Similar conclusions are found by independent study of \citep{HM2012}. 
If one assumes that the expansion history of the universe is given by 
the concordance $\Lambda$CDM model, then one plausible
explanation of this discrepancy is that the abundance of structure in the real 
Universe may be greater than expected, i.e.,
there might be more clusters and super clusters, and voids might have
grown larger and deeper. This explanation seems
to coincide with the first tension mentioned above. Again, one could
perhaps use this data to constrain non-Gaussianity,
and find a large $f_{\rm NL}$, but an alternative solution might be to modify gravity.

In this work, we explore the difference of structure formation in
GR and chameleon models of MG. Using the spherical collapse
model and excursion set theory \cite{bcek}, we investigate individual void properties and the void volume distribution
function in these two models.  We also compare the relative merits of distinguishing
between GR and MG using voids or halos; predictions for the latter have been addressed by \citet{le2012}.

One common way to distinguish MG from GR is by looking at the
difference between the lensing mass and dynamical mass of halos \citep{Feix2008, s2010, Jain2010, Zhao2010, cjk2012, ln2012}.
The chameleon model studied here predicts that such a difference is at most $1/3$
between the screened and unscreened cases. At present, it is still very
difficult to have mass estimates of halos which achieve this level of
accuracy, partly due to the difficulty of looking
for unscreened objects. To realize the $1/3$ difference, such objects must be both small, so that they are
not self-screened, and located in low
density environments, so as not to be screened by the environment.
Voids, however, are usually very low in density so that the fifth force is 
unscreened inside. Furthermore, we show that the strength of the fifth force may be relatively stronger than that of Newtonian gravity in voids.
This may lead to a larger difference of void properties from GR than that of halos.

The outline of this paper is as following: In section \ref{sect:chameleon}, we give a brief summary of the coupled scalar field
gravity, of which the chameleon model is an example. In section \ref{sect:static}, we solve the scalar field profile for
voids in this model and highlight interesting differences of the fifth
force to Newtonian gravity in voids. In section \ref{sect:single}, we extend the
spherical collapse model 
to solve for the evolution of shells in voids
in this model and identify the best regimes to distinguish this model
from GR. Section \ref{sect:method} presents the first crossing barrier for voids, and
incorporates the moving barrier and environmental dependence of
void formation to the excursion set theory to calculate a void
volume distribution function. We summarize our results and consider possible ways to test 
MG in voids in section \ref{sect:conclusion}.

\section{The Chameleon Theory}
\label{sect:chameleon}

This section lays down the theoretical framework for investigating the effects of a coupled scalar field in cosmology.  We shall present the relevant general field equations in \S~\ref{subsect:equations}, and then
specify the models analyzed in this paper  in \S~\ref{subsect:specification}.

\subsection{Cosmology with a Coupled Scalar Field}
\label{subsect:equations}

The equations presented in this subsection can be found in \citet{lz2009, lz2010, lb2011}, and  are presented here only to make this work self-contained.

We start from a Lagrangian density
\begin{eqnarray}\label{eq:lagrangian}
{\cal L} =
{1\over{2}}\left[\mpl^{2}{R}-\nabla^{a}\phi\nabla_{a}\phi\right]
+V(\phi) - C(\phi) ({\cal L}_{\rm{DM}} + {\cal L}_{\rm{S}}),
\end{eqnarray}
in which $R$ is the Ricci scalar; the reduced Planck mass is $\mpl=1/\sqrt{8\pi G}$ with $G$ being the
gravitational constant; and ${\cal L}_{\rm{DM}}$ and
${\cal L}_{\rm{S}}$ are respectively the Lagrangian
densities for dark matter and standard model fields. $\phi$ is
the scalar field and $V(\phi)$ its potential; the coupling
function $C(\phi)$ characterises the coupling between $\phi$
and matter. Given the functional forms for $V(\phi)$ and $C(\phi)$,
a coupled scalar field model is then fully specified.

Varying the total action with respect to the metric $g_{ab}$, we
obtain the following expression for the total energy momentum
tensor in this model:
\begin{eqnarray}\label{eq:emt}
T_{ab} = \nabla_a\phi\nabla_b\phi -
g_{ab}\left[{1\over2}\nabla^{c}\nabla_{c}\phi-V(\phi)\right]+ C(\phi) (T^{\rm{DM}}_{ab} + T^{\rm{S}}_{ab}),
\end{eqnarray}
where $T^{\rm{DM}}_{ab}$ and $T^{\rm{S}}_{ab}$ are the
energy momentum tensors for (uncoupled) dark matter and standard
model fields. The existence of the scalar field and its coupling
change the form of the energy momentum tensor, leading to
potential changes in the background cosmology and 
structure formation.

The coupling to a scalar field produces a direct
interaction (fifth force) between matter
particles due to the exchange of scalar quanta. This is best
illustrated by the geodesic equation for dark matter particles
\begin{eqnarray}\label{eq:geodesic}
{{{\rm d}^{2}\bf{r}}\over{{\rm d}t^2}} = -\vec{\nabla}\Phi -
{{C_\phi(\phi)}\over{C(\phi)}}\vec{\nabla}\phi,
\end{eqnarray}
where $\bf{r}$ is the position vector, $t$ the (physical) time, $\Phi$
the Newtonian potential and $\vec{\nabla}$ is the spatial
derivative; $C_\phi\equiv {\rm d}C/{\rm d}\phi$. The second term on the
right hand side is the fifth force, with potential $\ln C(\phi)$.

To solve the above two equations we need to know both the time
evolution and the spatial distribution of $\phi$, {\it i.e.}  we
need the solutions to the scalar field equation of motion (EOM)
\begin{eqnarray}
\nabla^{a}\nabla_a\phi + {{\rm{d}V(\phi)}\over{\rm{d}\phi}} +
\rho {\frac{\rm{d}C(\phi)}{\rm{d}\phi}} = 0,
\end{eqnarray}
where $\rho = \rho_{\rm DM} + \rho_{b}$, the sum of dark and
baryonic matter densities. Equivalently
\begin{eqnarray} \label{eq:eom0}
\nabla^{a}\nabla_a\phi + {{\rm{d}V_{\rm eff}(\phi)}\over{\rm{d}\phi}} =
0,
\end{eqnarray}
where we have defined
\begin{eqnarray}
V_{\rm eff}(\phi) = V(\phi) + \rho \, C(\phi).
\end{eqnarray}
The background evolution of $\phi$ can be solved easily given
the present-day value of  $\rho$ since 
$\rho \propto a^{-3}$. We can then divide $\phi$
into two parts, $\phi=\bar{\phi}+\delta\phi$, where
$\bar{\phi}$ is the background value and $\delta\phi$ is its
(not necessarily small nor linear) perturbation, and subtract the
background part of the scalar field equation of motion from the full equation
to obtain the equation of motion for $\delta\phi$. In the
quasi-static limit in which we can neglect time derivatives of
$\delta \phi$ as compared with its spatial derivatives (which
turns out to be a good approximation on galactic and cluster scales),
we find
\begin{eqnarray}\label{eq:scalar_eom}
\vec{\nabla}^{2} \delta \phi =
{{\rm{d}C(\phi)}\over{\rm{d}\phi}}\rho -
{{\rm{d}C(\bar{\phi})}\over{\rm{d}\bar{\phi}}}\bar{\rho} +
{{\rm{d}V(\phi)}\over{\rm{d}\phi}} -
{{\rm{d}V(\bar{\phi})}\over{\rm{d}\bar{\phi}}},
\end{eqnarray}
where $\bar{\rho}$ is the background matter density.

The computation of the scalar field $\phi$ using the above equation then completes the computation of the source term for the Poisson equation
\begin{eqnarray}\label{eq:poisson}
\vec{\nabla}^{2} \Phi &=& \frac{1}{2 \mpl^2}\left[\rho_{\rm tot}+3p_{\rm tot}\right]\nonumber\\
&=& {{1}\over{2 \mpl^2}}\left[\rho \, C(\phi) - 2V(\phi)\right], 
\end{eqnarray}
where we have neglected the kinetic energy of the scalar field because it is always very small for the model studied here.

\subsection{Specification of Model}
\label{subsect:specification}

As mentioned above, to fully fix a model we need to specify the functional forms of $V(\phi)$ and $C(\phi)$. Here we will use the models investigated by \cite{lz2009, lz2010, li2011}, with
\begin{eqnarray}\label{eq:coupling}
C(\phi) = \exp(\gamma \phi / \mpl),
\end{eqnarray}
and 
\begin{eqnarray}\label{eq:pot_chameleon}
V(\phi) = {{\rho_{\Lambda}}\over{\left[1-\exp\left(-\phi / \mpl \right)\right]^\alpha}}.
\end{eqnarray}
In the above $\rho_{\Lambda}$ is a parameter of mass dimension four and 
is of order the present dark energy density 
($\phi$ plays the role of dark energy in this model). $\gamma, \alpha$ are dimensionless  parameters controlling the strength of the 
coupling and the steepness of the potential respectively. 

We choose $\alpha\ll1$ and $\gamma>0$ as in \citet{lz2009, lz2010},
which ensure that $V_{\rm eff}(\phi)$ has a global minimum close to $\phi=0$
and that ${\rm d}^2V_{\rm eff}(\phi)/{\rm d}\phi^2 \equiv m^2_{\phi}$ at this
minimum is very large in high density regions. There are two
consequences of these choices of model parameters: (1) $\phi$ is
trapped close to zero throughout  cosmic history so that
$V(\phi)\sim \rho_{\Lambda}$ behaves as a cosmological constant; 
(2) the fifth force is strongly
suppressed in high density regions where $\phi$ acquires a large
mass, $m^2_{\phi}\gg H^2$ ($H$ is the Hubble expansion rate),
and thus the fifth force cannot propagate far. The suppression of the
fifth force is even stronger at early times, and thus its influence on
structure formation occurs mainly at late times.
The environment-dependent behaviour of the scalar field was
first investigated by \citet{kw}, and is often referred
to as the `chameleon effect'.

\section{Static underdensity solutions}
\label{sect:static}

The radial profile of a chameleon-type scalar field has been studied in detail for spherical overdensities, in which cases a simple analytical formula for the fifth force has been derived \citep{kw} and shown to agree well with the numerical simulations \citep{lzk2012}. We know from these previous studies that, depending on its size and environment, a spherical overdensity could develop a thin shell which is a region of fast change of $\phi(r)$ with respect to $r$, and approximately only the matter contained in this shell contributes to the fifth force on a particle at the edge of the overdensity. If the shell is thin the fifth force is much weaker than gravity (the latter coming from all mass contained in the overdensity), while if its thickness becomes comparable to the radius of the overdensity, the fifth force approaches a constant ratio to gravity. For our fiducial model this ratio is $2\gamma^2$ and we choose the coupling $\gamma$ such that $2\gamma^2 = 1/3$, so that the maximum deviations from GR match those of $f(R)$ models.

Unfortunately, no analytical approximation for the fifth force is known for the case of underdensities. It is our task in this section to study $\phi(r)$ in underdensities and the fifth force which results. We will see that the maximum ratio of $2\gamma^2 = 1/3$ will no longer apply in this case: in voids the fifth force can have much stronger effects than gravity.

\subsection{Voids in Newtonian Gravity}

Consider a spherically-symmetric underdensity defined by radius $r$ and inner and outer densities, $\rhoin$ and $\rhoout$, such that $\rhoin < \rhoout$. First we review the forces around such voids in Newtonian gravity. Since $C(\phi) \approx 1$, the first term on the right-hand side of Eq.~(\ref{eq:poisson})
can be integrated once to give the force per unit test mass
\be
F_{\rm N} (\chi) = -\frac{G M(< \chi)}{\chi^2}
\ee
where
\be
M (< \chi) = 4\pi \int_0^\chi \de \chi' \, \chi'^2 \rho_0(\chi') \, .
\ee
We are interested in the simplest model of a void, with top-hat density profile
\be \label{eq:density}
\rho_0 (\chi) = \left\{
\begin{matrix}
\rhoin \qquad {\rm for} \qquad \chi \le r \cr
\rhoout \qquad {\rm for} \qquad \chi > r
\end{matrix}
\right. \, .
\ee
(We use the notation $r$ for the void radius and $\chi$ for the radial coordinate for the sake of continuity with later sections of the paper.) The resulting force on the mass shell at $r$ is
\bea
F_{\rm N} (r) & = & - \frac{4\pi G}{3} \rhoin r \\
 & = & - \frac{\rhoin r}{6 \mpl^2} \, . \label{eq:F_N}
\eea
Only mass within the radius $r$ contributes to the force on it -- test masses inside completely empty voids where $\rhoin = 0$ feel no force since the pull from all the mass elements outside the void cancel perfectly. This is a standard, although counter-intuitive, result of Newtonian gravity. If $\rhoin$ is nonzero, the force on the shell is equal to that of a point particle of mass $M(< \chi)$ which is located at $\chi=0$, and the force is attractive.

Similarly, since $V(\phi) \approx \rho_{\Lambda}$, the second term on the right-hand side of Eq.~(\ref{eq:poisson}) gives the effective force due to the scalar field 
potential (or equivalently, the cosmological constant),
\be \label{eq:F_lam}
F_{\rm \Lambda} (r) = \frac{\rho_{\Lambda} r}{3 \mpl^2}.
\ee
This contributes an effective repulsive force at late cosmological times, which we call the dark energy force in this paper.

\subsection{Voids in Chameleon Theories}

The total force on a test particle is the sum of the Newtonian force, effective force from the dark energy, and the scalar-mediated fifth force. We will see that the fifth force is always {\it repulsive} in voids, in the sense that the force on a test mass pushes it away from the center of the void, towards the nearest wall; it aids the dark energy in emptying the void of matter.

The second term on the right-hand side of Eq.~(\ref{eq:geodesic}) gives the fifth force, which for our choice of $C(\phi)$ in Eq.~(\ref{eq:coupling}) is
\be \label{eq:F_5}
F_5 (\chi) = -\gamma \frac{\de}{\de \chi} (\phi / \mpl) \, .
\ee
We define the ratio of fifth to Newtonian forces as
\bea
\eta & \equiv & \frac{F_5}{F_{\rm N}} \ =\ \frac{6 \gamma \mpl}{r \rhoin} \left. \frac{\de \phi}{\de \chi}\right|_{\chi = r} \, , \label{eq:eta}
\eea
which is constrained to be $\eta \le 2\gamma^2$ for overdensities. Thus the problem of finding the force deviations on a test particle in the void has been reduced to obtaining the scalar field profile $\phi (\chi)$. Before solving Eq.~(\ref{eq:eom0}) to obtain the profile, we note some properties of this scalar field model which will simplify the solution.

At fixed density $\rho_0$, our theory has an effective potential
\be \label{eq:eff}
V_{\rm eff} (\phi) = {{\Lambda}\over{\left[1-\exp\left(-\phi / \mpl \right)\right]^\alpha}} + \rho_0 \exp(\gamma \phi / \mpl).
\ee
Call $\phi_0$ the field value which minimizes this potential for the given density. Using the facts that $\alpha \ll 1$ and $\phi_0 / \mpl \ll 1$ (Sec.~\ref{subsect:specification}), we set $\partial V_{\rm eff} / \partial \phi = 0$ and expand in the small parameter $\phi_0 / \mpl$ to find
\be
\phi_0 / \mpl = \frac{\alpha}{\gamma} \frac{\rho_\Lambda}{\rho_0} \, .
\ee
If $m_0$ is the mass of small fluctuations about this minimum, then
\be
m_0^2 \ = \ \frac{\partial^2 V_{\rm eff}}{\partial \phi^2} = \frac{(\gamma \rho_0)^2}{\alpha \mpl^2 \rho_\Lambda}
\ee
so that the associated Compton wavelength $\lambda_0 \equiv m_0^{-1}$ and the field value at the minimum are related by
\be
\phi_0 = \sqrt{\alpha \rho_\Lambda} \lambda_0 \, .
\label{eq:lam}
\ee
The above analytic relations between the density and associated field value and compton wavelength, namely $1/\rho_0 \propto \phi_0 \propto \lambda_0$, are not a general feature of scalar-tensor theories of gravity, nor even of chameleon models. For example, in the $f(R)$ model of \citet{hs2007}, the relation between these three quantities has no closed form solution. While these analytic relations are useful in themselves, we now show how they can be used to decrease the void parameter space from three to two variables, while simultaneously removing dependence on the theory parameters $\alpha$ and $\gamma$.

Naively, any top-hat void of radius $r$ and density $\rhoin$ in a uniform background density $\rhoout$ is dependent on three length scales: $r$, $\lamin$, and $\lamout$. However we show that since $\phi / \mpl \ll 1$, the Planck scale drops out of the equation of motion, giving us the freedom to rescale the solution by one of these lengths. This reduces the problem to two non-trivial degrees of freedom. 
The equation of motion Eq.~(\ref{eq:eom0}) is given by
\be \label{eq:eom1}
\frac{\de^2\phi}{\de \chi^2} + \frac{2}{\chi}\frac{\de \phi}{\de \chi} = -\alpha \frac{\rho_\Lambda \sqrt{\kappa} e^{-\sqrt{\kappa} \phi}}{(1-e^{-\sqrt{\kappa}\phi})^{\alpha+1}} + \rho_0 (\chi) \, \gamma \sqrt{\kappa} \, e^{\gamma \sqrt{\kappa} \phi} \, ,
\ee
where $\rho_0 (\chi)$ is again the top-hat profile of Eq.~(\ref{eq:density}).
Expanding to lowest order in $\phi / \mpl$ and using $\alpha \ll 1$ we have
\bea
\frac{\de^2 \phi}{\de \chi^2} + \frac{2}{\chi}\frac{\de \phi}{\de \chi} & = & \alpha \rho_\Lambda \left(\frac{1}{\phi_0 (\chi)} - \frac{1}{\phi} \right) \\
 & = & \frac{\alpha \rho_\Lambda}{\phiout} \left(\frac{\phiout}{\phi_0 (\chi)} - \frac{\phiout}{\phi} \right) \, .
\eea
Defining the dimensionless field $\psi \equiv \phi / \phiout$ and using (\ref{eq:lam}) yields
\be
\frac{\de^2 \psi}{d\chi^2} + \frac{2}{\chi}\frac{\de \psi}{\de \chi} = \frac{1}{\lamout^2} \left(\frac{\phiout}{\phi_0 (\chi)} - \frac{1}{\psi} \right) \, .
\ee
Then defining a dimensionless radial coordinate $\tau \equiv \chi / \lamout$, the equation further simplifies to
\be
\frac{\de^2 \psi}{\de \tau^2} + \frac{2}{\tau}\frac{\de \psi}{\de \tau} = \frac{\phiout}{\phi_0 (\tau)} - \frac{1}{\psi}\, .
\ee
Now, from the three length scales we can form two ratios $ r / \lamout$ and $\lamout / \lamin$ (note that for voids we must have $0 \le \lamout / \lamin < 1$) and recast the EOM in terms of these. The problem is then reduced to solution of the differential equation
\be \label{eq:eom}
\frac{\de^2 \psi}{\de \tau^2} + \frac{2}{\tau}\frac{\de \psi}{\de \tau} + \frac{1}{\psi} = \left\{
\begin{matrix}
\lamout / \lamin \qquad {\rm for} \qquad \tau \le r / \lamout \cr
1 \qquad {\rm for} \qquad \tau > r / \lamout
\end{matrix}
\right.
\ee
with boundary conditions
\be \label{eq:bc}
\left.\frac{\de \psi}{\de \tau}\right|_{\tau = 0} = 0, \qquad \qquad \psi(\tau \rightarrow \infty) = 1 \, .
\ee
Rewriting Eq.~(\ref{eq:eta}) in terms of the new variables and using $\rho_\Lambda = \Omega_{\Lambda} \rho_{\rm c}$, we find
\be
\eta (r) = 6 \gamma \sqrt{\frac{\alpha \Omega_{\Lambda}}{\Omega_m}} \frac{\mpl \sqrt{\bar{\rho}_m}}{r \rhoin}  \left. \frac{\de \psi}{\de \tau}\right|_{\tau = r / \lamout} \, ,
\ee
where $\bar{\rho}_m$ is the background matter density today.

Before describing the resulting solutions of the scalar field and fifth force for realistic underdensities, we make some comments about the relevance of Eq.~(\ref{eq:eom}) for our results in Sections~\ref{sect:single} and~\ref{sect:method}. Since there is no known analytical approximation for $\phi$ in underdensities, as there is in the overdense case, it will be necessary to solve numerically the EOM at each time step for an expanding void in Sec.~\ref{sect:single}. Furthermore, in order to obtain the void-formation barriers of Sec.~\ref{sect:method}, we must calculate the trajectories of many such expanding voids of different initial sizes and densities. While a top-hat underdensity intrinsically has three degrees of freedom, $\rhoin$, $\rhoout$, and $r$, we have shown that two ratios formed from these quantities are sufficient to solve the EOM. Thus, the most difficult numerical challenge of Sections~\ref{sect:single} and~\ref{sect:method} can be overcome with a single two-dimensional table of $\de \psi / \de \tau$ values, where the derivative is evaluated at the border of the void. Furthermore, our recasting of Eq.~(\ref{eq:eom1}) as Eq.~(\ref{eq:eom}) has no explicite dependence on the theory parameters $\alpha$ and $\gamma$. Thus, this same 2-D table serves to calculate the void-formation barriers under variations in $\alpha$ and $\gamma$, as in Sec.~\ref{sect:vary-ag}.

\subsection{Radial Profile of the Scalar Field}
\label{sect:scalar}

\begin{figure*}
\centering
\resizebox{180mm}{!}{\includegraphics{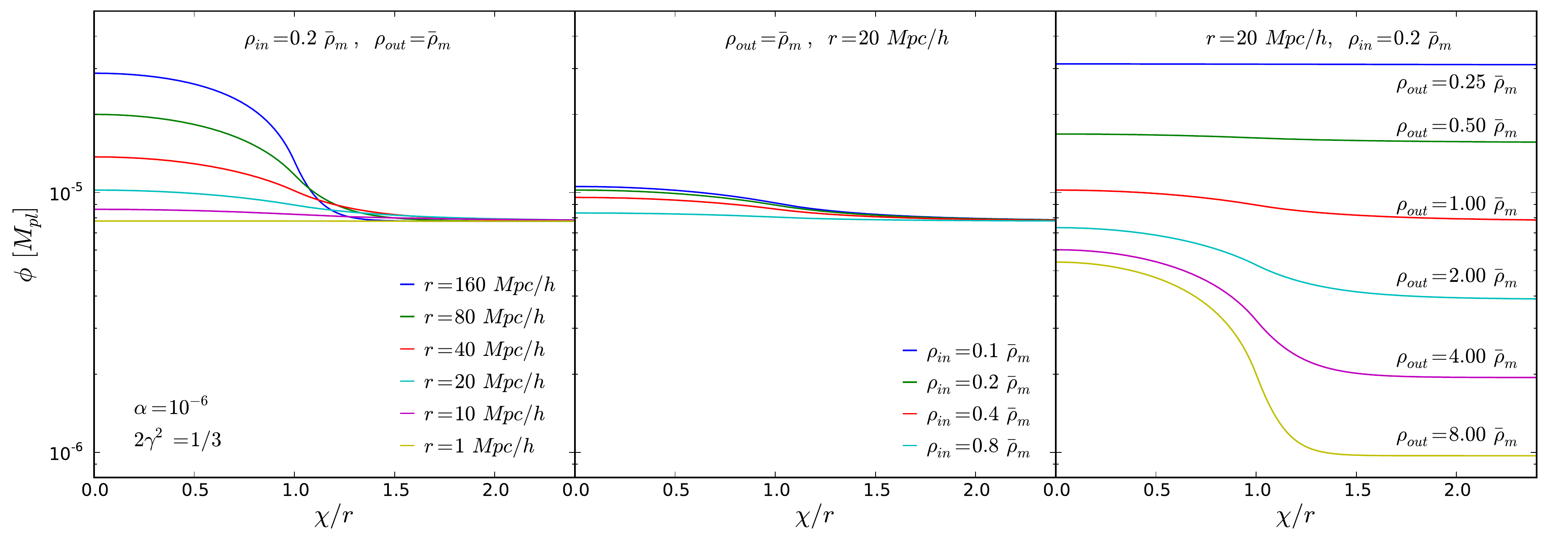}}
\caption{{\it Left panel}: Radial profile of the scalar field in a spherical top-hat underdensity for fixed values of $\rhoin = 0.2 \, \bar{\rho}_m$, $\rhoout = \bar{\rho}_m$ and different radii $r$. {\it Center panel}: The same, but for fixed values of $\rhoout = \bar{\rho}_m$, $r = 20 \, {\rm Mpc}/h$ and different inner densities $\rhoin$. {\it Right panel}: The same, but for fixed values of $r = 20 \, {\rm Mpc}/h$, $\rhoin = 0.2 \, \bar{\rho}_m$ and different outer densities $\rhoout$. Note that the horizontal axis is scaled with respect to void radius $r$, so $\chi / r = 1$ is the edge of the spherical underdensity; also we evaluate the cosmic mean density at the present day, $\bar{\rho}_m (z) = \bar{\rho}_m (0)$.}
\label{fig:scalar}
\end{figure*}

Now we consider the results for the radial profile of the scalar field in various underdensities, paying special attention to the value of the derivative at the void border. 
The left panel of Fig.~\ref{fig:scalar} shows the dependence on void radius, $r$. If $r$ is small, then the underdensity can be considered as a small perturbation on the environment and the scalar field value inside is very close to its value at the boundary. As $r$ increases, however, there is increasing space for $\phi$ to evolve away from the exterior value (here $\phiout \approx 0.8 \times 10^{-5} \mpl$) as $\chi/r$ decreases, and therefore the scalar comes closer to reaching the value which minimizes the interior effective potential. Since $\phi_{\rm in (out)} \propto 1/\rho_{\rm in (out)}$ and $\rhoout / \rhoin = 5$ in the figure, we know $\phiin = 5 \phiout$ and see that even $160 \, {\rm Mpc}/h$ is not enough space for the scalar field to attain its minimum at the center of the void. 

Fig.~\ref{fig:scalar} shows the dependence of the scalar field profile $\phi(r)$ on interior density $\rhoin$ in the central panel, assuming an exterior density equal to the cosmic mean. Here the field does not experience much change between the outside and inside of the void, growing by only 25\% in the most extreme case, $\rhoin = 0.1 \, \bar{\rho}_m$. As a result, the derivative of the scalar, and therefore the fifth force, at the void border $\chi / r = 1$ must be small. However, we will see that in order to get a full picture of the forces involved it is necessary to consider the gravitational force and dark energy force as well. For this void the magnitude of the fifth force is about twice as large as Newtonian gravity, so that even this slowly varying $\phi$ profile results in a force that is stronger than $F_{\rm N}$.

Finally, the dependence on $\rhoout$ is shown in the right panel of Fig.~\ref{fig:scalar}. The variations here appear more drastic, since only in this panel is the limiting value $\phiout$ changed from one curve to another. With fixed interior density, a denser environment for the void results in a larger change in the scalar and correspondingly higher derivative $\de \phi / \de \chi$. Note also that due to Birkhoff's theorem, changes in $\rhoout$ do not affect the gravitational force inside the void, nor is the dark energy force is affected. So only from this panel can we infer directly that larger gradients of $\phi$ imply greater deviations from GR.

There are some interesting differences from the overdense case. Consider an overdensity and underdensity each embedded in the same environmental density $\rhoout'$, with corresponding minimum $\phiout'$. For the overdensity, we know $\phi$ decreases from $\phiout'$ as we move towards the center; however by the shape of the effective potential Eq.~(\ref{eq:eff}), $\phi$ is strictly positive, so $0 < \phiin' < \phiout'$. The maximum change is therefore $\Delta \phi = \phiout'$, no matter how great is the interior density. In contrast, for the underdensity, $\phi$ increases from $\phiout'$ as we move towards the center so that $\Delta \phi$ has no such bound: $\rho_{in}$ can be infinitely small in principle. For concreteness consider the lowest curve on the right panel of Fig.~\ref{fig:scalar}: here $\phiout' = 10^{-6} \mpl$ so that for an overdensity $\Delta \phi < 10^{-6} \mpl$, while for the pictured underdensity we see $\Delta \phi \ge 4 \times 10^{-6} \mpl$. Since the fifth force is proportional to the derivative of $\phi$ at the void border, we expect this lack of upper bound on $\Delta \phi$ for underdensities to show itself in the force. We turn our attention next to these results.

\subsection{The Fifth Force}
\label{sect:fifth_force}

\begin{figure*}
\centering
\resizebox{180mm}{!}{\includegraphics{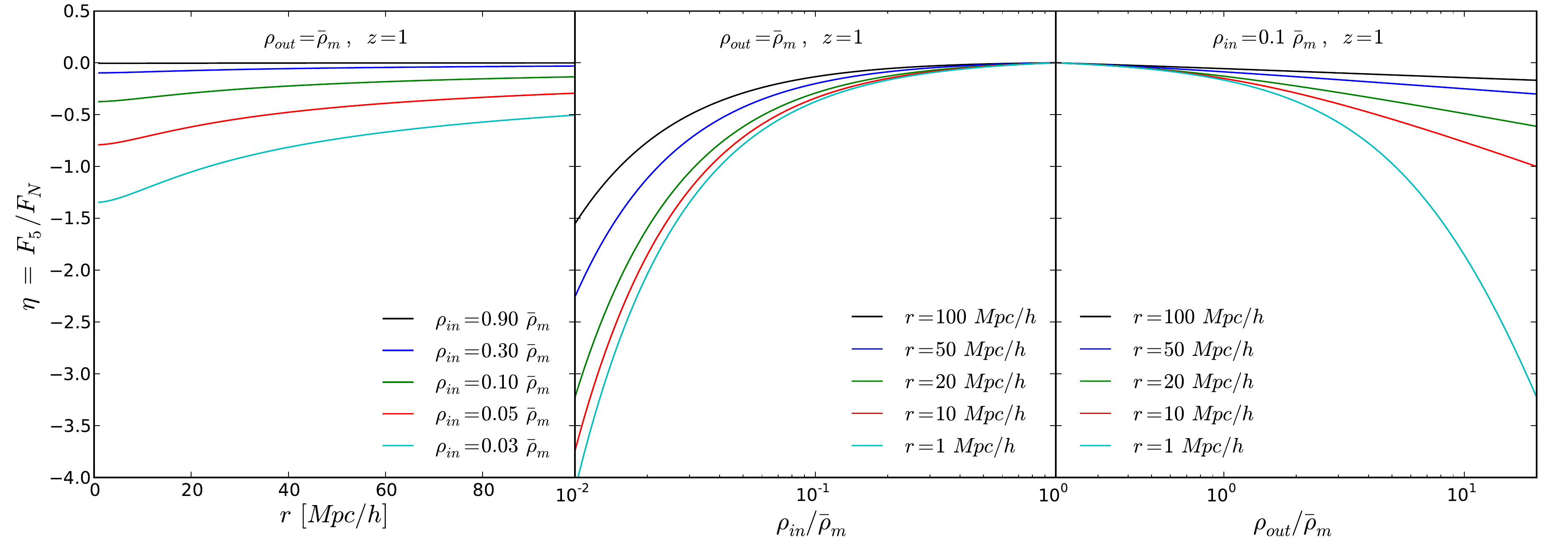}}
\resizebox{180mm}{!}{\includegraphics{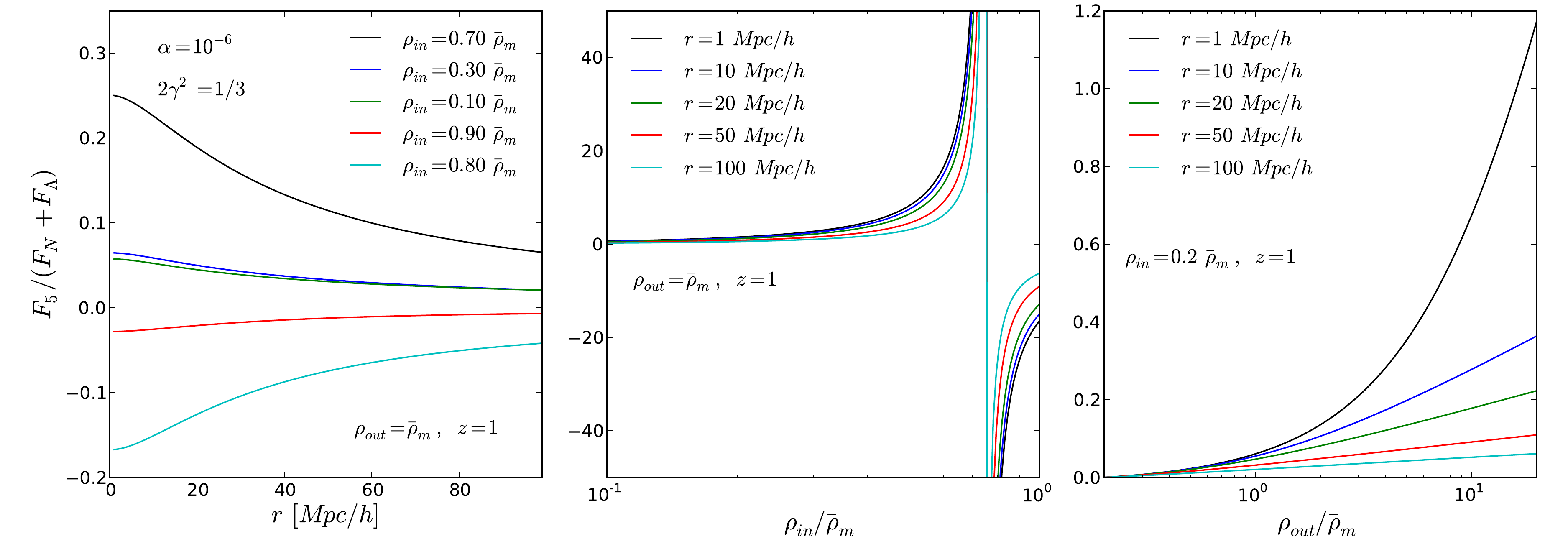}}
\caption{
{\it Top-left panel:} Variations of the force deviation $\eta$ with underdensity radius, $r$. The exterior density is fixed to the cosmic mean today, $\rhoout = \bar{\rho}_m$. 
Various values of interior density $\rhoin$ are shown, with $\rhoin$ decreasing from top to bottom.
{\it Top-center panel:} The same, but for continuous variations of $\rhoin$, fixed $\rhoout = \bar{\rho}_m$ and various values of radius $r$, with $r$ decreasing from top to bottom.
{\it Top-right panel:} The same, but for continuous variations of $\rhoout$, fixed $\rhoin = 0.1 \, \bar{\rho}_m$, and various values of radius $r$, with $r$ decreasing from top to bottom.
Note that in all the panels, we evaluate the cosmic mean density at redshift one, $\bar{\rho}_m (z) = \bar{\rho}_m (1)$.
{\it Bottom-left, -middle} and {\it -right panels} are the same as the top-left, -middle and -right panels, but showing the fractional difference 
of the total force between MG and GR theories, $F_5$/($F_N$+$F_{\Lambda}$).}
\label{fig:eta}
\end{figure*}

In the top panels of Fig.~\ref{fig:eta} we show the force deviation $\eta = F_5 / F_{\rm N}$ with variations in the three physical parameters which define a void, $r$, $\rhoin$, and $\rhoout$. The first interesting feature is that $\eta$ is always negative. The fifth force in voids is repulsive, always pointing at the opposite 
direction of normal gravity. This is the direct consequence of the scalar field profile we have shown in Fig.~\ref{fig:scalar}, whose slope is always negative at the edge of the underdensity. Intuitively, this repulsion occurs due to the Yukawa potential ($e^{-\chi/\lambda}/\chi$) of the scalar: at distances of the order of the Compton wavelength, the potential falls off more strongly than $1/\chi$. Mass elements on the far wall of a large void are unable to cancel the pull of the near wall. Furthermore, even if $\rhoin$ is nonzero, the integrated mass inside the shell is unable to compete with the denser nearby wall, and the force is again repulsive.

Secondly, as we anticipated in Sec.~\ref{sect:scalar}, the unboundedness of the field, along with the result of Eq.~(\ref{eq:F_N}) that $F_{\rm N}$ vanishes as $\rhoin \rightarrow 0$ or $r \rightarrow 0$, leads to deviations which do not share the bound of $|\eta| \le 2 \gamma^2$. Thus the relative strength of the fifth force can be much larger than Newtonian gravity, as seen in all top panels of Fig.~\ref{fig:eta}. Even for common voids with ratio of densities $\rhoin / \rhoout = 0.2$ we can have $|\eta| \approx 1/3$, already reaching the upper bound for overdensities. If the ratio decreases to the percent level, then $\eta \sim -2$ for the smallest voids. 

The left panel of Figs.~\ref{fig:scalar} and top-left panel of Fig.~\ref{fig:eta} both show variations with respect to void radius $r$. Comparing these, we see that while the change in $\phi$ and therefore the fifth force increases with void radius, the deviation $\eta$ gets smaller. Thus we infer that $F_{\rm N}$ increases more quickly than $F_5$ in these cases due to the increasing mass enclosed within the larger void radius.

In contrast, comparing the middle panel of Figs.~\ref{fig:scalar} and top-middle panel of \ref{fig:eta} shows that under variations in $\rhoin$ the changes in the fifth force dominate the dependence of $\eta$. The net effect of decreasing the interior density is to strengthen the fifth force relative to gravity.

The variations of $\rhoout$ in the top-right panel of Fig.~\ref{fig:eta} leave $F_{\rm N}$ unaffected, so here changes in $\eta$ straightforwardly reflect changes in $F_5$. We can unify the results of varying $\rhoout$ and $\rhoin$ by noting that increasing the density contrast $\rhoout / \rhoin$ generally increases the deviation from GR.

In principle this unboundedness of the force ratio $\eta$ in underdensities looks very promising for distinguishing between GR and chameleon models. However, at late times when $\bar \rho_m$ and $\rho_{\Lambda}$ are comparable, the repulsive dark energy force can dominate over Newtonian gravity where the density is low. $F_{\Lambda}$ is common in both GR and MG models but negligible for halos where the local density is much greater than the cosmic mean. The evolution of voids in MG models are therefore affected by $F_5$, $F_{\Lambda}$ and $F_N$. 

Bottom panels of Fig.~\ref{fig:eta} show the fractional difference of total force between MG and GR, $F_5$/($F_N$+$F_{\Lambda}$). Comparing them with the top 
panels, we find the following: A.) like $\eta$, the fractional difference decreases with radius (bottom-left) and increases with 
$\rhoout$(bottom-right). This is because the additional F$_{\Lambda}$ term is just a constant at a certain epoch. B.) $F_5$/($F_N$+$F_{\Lambda}$) can be positive or negative, depending on the relative amplitude of $F_N$ and $F_{\Lambda}$. The transition occurs at $\rhoin=2\rho_{\Lambda}$ when $F_N$ is canceled out by $F_{\Lambda}$, and the evolution of the system is only governed by $F_5$. Note that the sign switch in $F_5$/($F_N$+$F_{\Lambda}$) is no more than an indicator for the switch of the relative strength between $F_N$ and $F_{\Lambda}$. The forces $F_{\Lambda}$ and $F_5$ are always repulsive, and act to accelerate the expansion of void.  C.) When $\rhoin$ is close to $2\rho_{\Lambda}$, the fractional difference can be very large. 

In summary, if we track the evolution of a spherical underdensity with the radius of $r$, in the early universe it is dominated by $F_N$, the amplitude of which decreases with $\rho_{in}$. Later, the repulsive dark energy force $F_{\Lambda}$ from the background scalar field emerges to cancel part of $F_N$, and it helps to accelerate the expansion of void shells. In the mean time, $F_5$ appears from the coupling of the scalar field with mass, and is also repulsive in voids. As the void keeps emptying itself, $F_5$ becomes larger and $F_{\Lambda}$ also grows with time as $\Omega_{\Lambda}$ increases. The amplitude of the positive $F_N$+$F_{\Lambda}$ keeps decreasing until $\rhoin = 2 \rho_{\Lambda}$, then $F_N + F_{\Lambda}$ switches sign and the amplitude starts increasing. $F_5$ should also keep increasing with time as $\rhoin$ decreases faster than its environment density, which makes the density contrast inside and and outside the void grow larger. Overall, $F_5$ should help to accelerate the expansion of void. In the next section, we will quantify this effect.

\section{Evolving individual void}
\label{sect:single}

With the solution of the fifth force in underdense regions, we can apply it to solve  the equations that govern the evolution of a spherical underdensity in a given environment specified by its density. We will explore how the evolution of voids or underdense regions are affected by the fifth force.

\subsection{Evolution of Environment}

We have shown in the previous section that the profile of the scalar field and hence the fifth force depends on the local density as well as the density of 
its environment. This is one distinct feature of chameleon models. We therefore need to follow the evolution of the environment properly in order to calculate the fifth force. The environmental dependence in chameleon models has been discussed by 
\citet{le2012} and \citet{ll2012} for halos. We shall adopt the same idea of taking the environment as a spherical region with radius much larger than the 
underdensity in consideration. The exact choice of the environment size will be specified where it is used later for the void statistics (Sec.~\ref{subsect:moving}). Note that for the purposes of single-shell evolution which we describe in this section, the environment is completely specified by its density relative to the cosmic mean.

To track the non-linear evolution of the environment, we denote its physical radius at time $t$ by $r(t)$, its initial comoving radius by $R$, and define $q(t)\equiv a(t)R$. 
The evolution equation for $r(t)$ is
\begin{eqnarray}\label{eq:revolution}
\frac{\ddot{r}}{r} &=& -\frac{1}{6\mpl^2}\left(\rho - 2\rho_\Lambda\right),
\end{eqnarray}
where $\rho \equiv 3M/4\pi r^3$ is the matter density in the spherical region of the environment and the constant $\rho_\Lambda\approx V(\phi)$ is the 
effective dark energy density.
Note that Eq.~(\ref{eq:revolution}) assumes that the environment is unaffected by the fifth force. We make this approximation since the environments are very large in size and therefore the effects of the fifth force on them are minimal.
Let us define $y(t)\equiv r(t)/q(t)$ and change the time variable to 
$N\equiv\ln(a)$; derivatives with respect to $N$ are denoted by $y' = \de y / \de N$. By using Eq.~(\ref{eq:revolution}), $q(t)\propto a(t)$ and the Friedman equation $H^2/H_0^2 = \Omega_m a^{-3} + \Omega_\Lambda$, we find
\begin{eqnarray}\label{eq:yevolution}
y'' + \left[2-\frac{3}{2}\Omega_m(N)\right]y' + \frac{\Omega_m(N)}{2}\left(y^{-3}-1\right)y &=& 0,
\end{eqnarray}
which is a non-linear equation, where $\Omega_m(N) \equiv \Omega_me^{-3N}/(\Omega_me^{-3N}+\Omega_\Lambda),$ and 
$\Omega_\Lambda(N) \equiv \Omega_\Lambda/(\Omega_me^{-3N}+\Omega_\Lambda)$.

At very early times we must have $y\approx1$ and so can write $y=1+\epsilon$ with $|\epsilon|\ll1$. Substituting this into Eq.~(\ref{eq:yevolution}) to get the linearised evolution equation for $\epsilon$, we find that $\epsilon\propto D_+$, in which $D_+$ is the linear growth factor governed by the equation
\begin{eqnarray}\label{eq:dplus}
D''_++\left[2-\frac{3}{2}\Omega_m(N)\right]D'_+-\frac{3}{2}\Omega_m(N)D_+ &=& 0,
\end{eqnarray}
and the proportionality coefficient can be found using mass conservation: $y^3(1+\delta_i)=1\Rightarrow \epsilon=-\delta_i/3\propto D_+$ (here $\delta_i$ is the linear density perturbation at the initial time). As a result, the initial conditions for $y$ are $y(a_i)=1-\delta_i/3$ and $y'(a_i)=-\delta_i/3$.

Eqs.~(\ref{eq:yevolution}, \ref{eq:dplus}), associated with their corresponding initial conditions, completely determine the necessary dynamics in the $\Lambda$CDM model used for the environment shell. In what follows we shall use $y_{\rm env}$ to denote the $y$ for the environment, in contrast to that for the underdensity, which we shall denote by $y_{\rm v}$. We will reserve $r$ for the physical radius of the underdensity, matching the notation of Sec.~\ref{sect:static}.

\subsection{Evolution of Underdensity}

The only difference between the evolution of an underdensity and that of its environment is the effect of the fifth force. To calculate the fifth force at each time-step we use a spherical top-hat profile,
\be \label{eq:density-evo}
\rho (\chi) = \left\{
\begin{matrix}
\rho_{\rm v} \qquad {\rm for} \qquad \chi \le r \cr
\rho_{\rm env} \qquad {\rm for} \qquad \chi > r
\end{matrix}
\right. \, .
\ee
We assume there is no shell 
crossing, so that to study the evolution we only need to understand the motion of the shell at the edge. Note that this is not strictly true: for a model different than ours, \citet{ms2009} have shown that modified gravity can cause an initially top-hat underdensity to have a slight density gradient near the edge. We find a similar effect, but it is quite small and it is beyond the scope of this paper to self-consistently track the deviations of the density profile from the top-hat.

Denoting the density inside the underdensity by $\rho_{\rm v}$ and using mass conservation, we can show that
\bea
\rho_{\rm v}r^3 & = & \left(\bar{\rho}_ma^{-3}\right)(aR)^3 \nonumber \\
\rho_{\rm v} & = & \bar{\rho}_m\left(ay_{\rm v}\right)^{-3} \, , \label{eq:rhov_yv}
\eea
where $\bar{\rho}_m$ is the background matter density today. Similarly, the matter density in the environment, $\rho_{\rm env}$, can be expressed in terms of $y_{\rm env}$ as
\begin{eqnarray}\label{eq:rhoenv_yenv}
\rho_{\rm env} = \bar{\rho}_m\left(ay_{\rm env}\right)^{-3}.
\end{eqnarray}
Using these relations we can rewrite Eqs.~(\ref{eq:F_N}) and~(\ref{eq:F_5}) in terms of the variables $y_{\rm v}$ and $y_{\rm env}$, yielding
\begin{eqnarray}
F_{\rm N} &=& \frac{1}{6\mpl^2}\bar{\rho}_m\left(ay_{\rm v}\right)^{-2}R\nonumber\\
&=& \frac{1}{2}\Omega_m\left(H_0R\right)\left(ay_{\rm v}\right)^{-2}H_0,\\
F_5 &=& \gamma\frac{{\rm d}\left(\phi / \mpl \right)}{{\rm d}\chi}\Big|_{\chi=r}\nonumber\\
&=& \sqrt{3\alpha\Omega_\Lambda}\gamma H_0\frac{{\rm d}\psi}{{\rm d}\tau}\Big|_{\tau = r / \lamout} \, ,
\end{eqnarray}
where
\be
r / \lamout = \sqrt{\frac{3}{\alpha\Omega_\Lambda}}ay_{\rm v}\left(ay_{\rm env}\right)^{-3}\gamma\Omega_mH_0R \, ,
\ee
and $\psi$ and $\tau$ are defined as in Sec.~\ref{sect:static}.
The fifth-force-to-gravity ratio is then
\begin{eqnarray} \label{eq:eta-evo}
\eta &=& \frac{\sqrt{3\alpha\Omega_\Lambda}\gamma\frac{{\rm d}\psi}{{\rm d}\tau}\Big|_{\tau = r / \lamout}}{\frac{1}{2}\Omega_m\left(H_0R\right)\left(ay_{\rm v}\right)^{-2}} \, ,
\end{eqnarray}
and the evolution equation of the underdensity becomes
\begin{eqnarray}\label{eq:revolution2}
\frac{\ddot{r}}{r} &=& -\frac{1}{6\mpl^2}\left[\rho_{\rm v} (1+\eta)-2\rho_\Lambda\right].
\end{eqnarray}
Rewriting using $y_{\rm v}$ we obtain
\begin{eqnarray}\label{eq:yevolution2}
y''_{\rm v} + \left[2-\frac{3}{2}\Omega_m(N)\right]y'_{\rm v} + \frac{\Omega_m(N)}{2} \, [y_{\rm v}^{-3} (1 + \eta) -1] y_{\rm v} = 0 \, .
\end{eqnarray}
Note that we absorb all the difference between GR and MG in $\eta$ in the above equation, which is the same quantity we have shown in 
the top panels of Fig.~\ref{fig:eta}.  Equations~(\ref{eq:yevolution}, \ref{eq:dplus}, \ref{eq:eta-evo}, \ref{eq:yevolution2}) form a set of coupled nonlinear differential equations, which govern the evolution of an 
underdensity in a given environment.

\begin{figure*}
\centering
\resizebox{180mm}{!}{\includegraphics{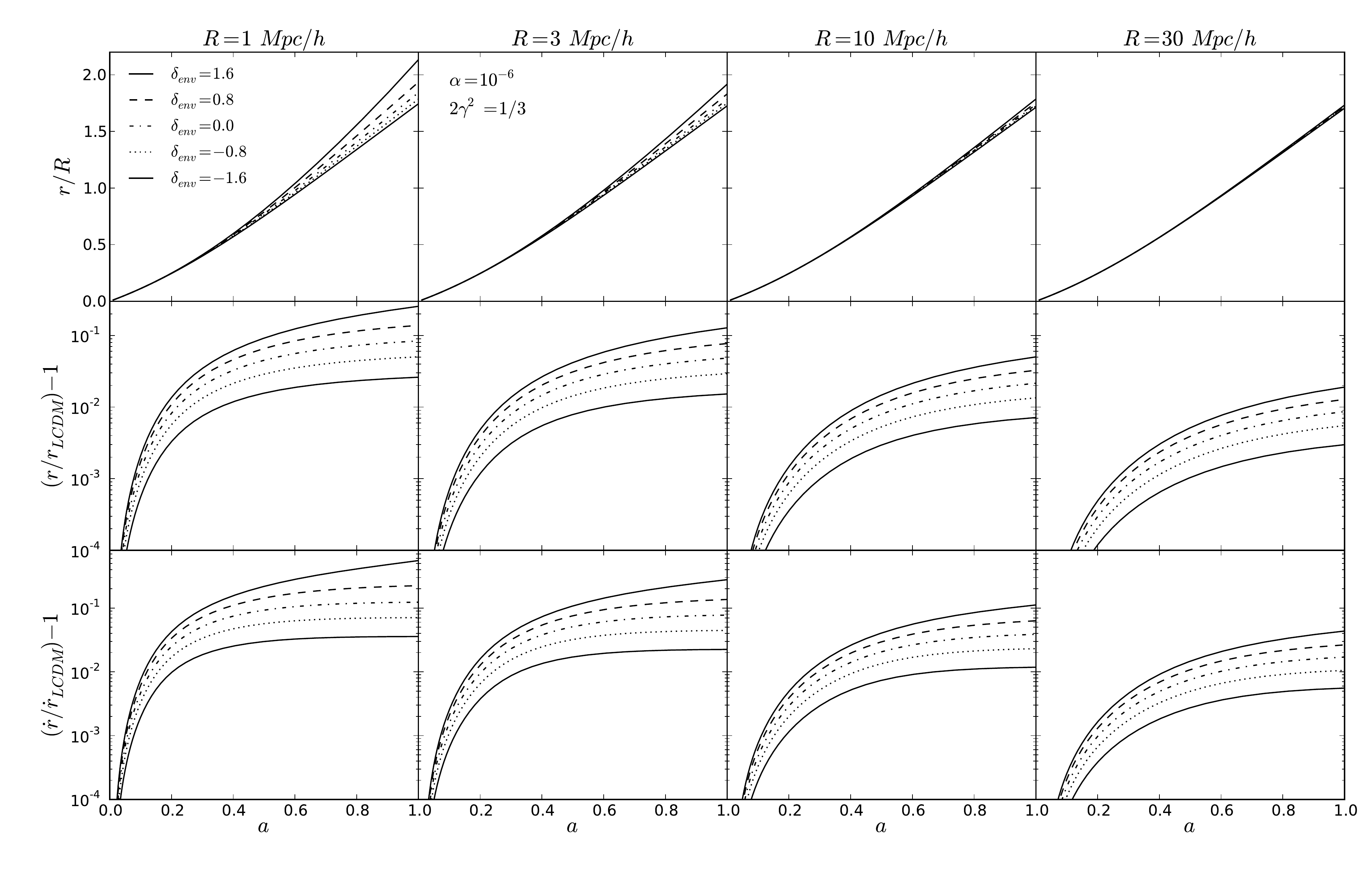}}
\caption{{\it Top row:} Radius $r$ of an expanding underdensity in units of its initial comoving radius $R$, as a function of scale factor $a$.
{\it Center row:} Fractional difference between the radii of such underdensities with identical initial conditions, expanding with and without the fifth force.
{\it Bottom row:} Fractional difference in the velocity.
Columns show various values of initial comoving radius, $R = 1, 3, 10$ and $30 \, {\rm Mpc}/h$, from left to right.
All panels have an initial underdensity, linearly extrapolated to today, of $\delta = -2.76$: these are objects which in a universe with no fifth force would have just reached the epoch of shell-crossing today. Various values of the exterior density are shown, with $\denv$ decreasing from top to bottom. The largest deviations from GR occur for voids expanding within a larger overdense region.}
\label{fig:single}
\end{figure*}

We can now solve the above equations to track the evolution of a spherical top-hat void. We compare results in our chameleon model and in $\Lambda$CDM in Fig.~\ref{fig:single}; both start from the same underdense regions $\dsc = -2.76$, where  $\dsc$ (shell-crossing) is the initial density contrast of the void region extrapolated to today. This setting of the initial condition corresponds to voids that would have just shell-crossed today in the
$\Lambda$CDM universe \citep{sv2004}. The mean nonlinear density contrast of those underdense regions today is $\delta = -0.8$, so that even without the fifth force these are already 
fairly empty voids.

The difference between the two models in the void expansion history depends on the initial comoving sizes of voids $R$ as well as their environment, quantified by $\delta_{\rm env}$, the initial 
environment density perturbation linearly extrapolated to today. Voids in denser environments show a larger difference between GR and MG. 
This is due to the greater density contrast realized by an underdensity in a very overdense environment. As seen in Figs.~\ref{fig:scalar} and~\ref{fig:eta},
such contrasts in density cause a large change in the scalar field, which in turn results in a stronger fifth force. 

In all cases, voids in MG expand faster and grow larger than their counterparts in 
$\Lambda$CDM. The comoving void radius would have grown by a factor of 1.7 at shell-crossing in GR. However in MG, Fig.~\ref{fig:single} shows the same underdensity would have grown by a factor of
$\sim$ 2 for voids with $R\sim 3$ Mpc/$h$ in dense environments. The difference between GR and MG is at $\sim 10\%$ level, and smaller for less dense 
environments. For larger voids the difference becomes smaller, e.g.,
for $R\sim$100 Mpc/$h$ it is at the sub-percent level. Although the absolute value of the fifth force is smaller for small voids (left panel of Fig.~\ref{fig:scalar}), the gravitational force is correspondingly smaller due to the decreased integrated mass. As shown in Fig.~\ref{fig:eta}, the net effect is that the instantaneous ratio between the two is larger 
(more negative) for smaller voids. The void size $y_v$ or $r$ at any given time shows the integrated effects of nonzero $\eta$ from all previous times. Thus the radii 
of smaller voids have expanded more beyond their $\Lambda$CDM counterparts, which themselves have expanded much more than the background. 

While the size of voids shows the cumulative effect of gravity, the expansion velocity of each shell responds more sensitively 
to any change of gravity at a given time. The bottom panels of Fig.~\ref{fig:single} shows the fractional difference of the expansion velocity of 
shells in GR and MG. Indeed, the differences in velocity are larger than the differences in sizes. For voids of $R \, \sim 3 $ Mpc/$h$, 
the expansion velocity can be $10\%$ to $30\%$ faster in MG in over-dense environments. By $R\sim 30$ Mpc/$h$, the difference has dropped to a few percent in this model.

Our results suggest that perhaps the best way to look for modified gravity is to find voids in overdense environments, especially small voids, where we 
expect the difference from GR is maximized. Those voids should be emptier due to the relatively strong repulsive fifth force and faster 
expansion of the shells. Moreover, the difference in redshift space could be 
more prominent due to the even larger difference in the velocity field. We propose that the clustering analysis of tracers of small voids in redshift space could be a powerful test of GR. Predictions for this test from $N$-body simulations will be presented in a separate paper.

\section{Void definition and statistics}
\label{sect:method}

Having success in following the evolution of a single shell, 
we can now look for a common definition of voids for GR and MG. Then we will 
compare the population of voids in both GR and MG statistically by generalizing 
the excursion set approach \citep{bcek}. 
But first we will lay down briefly the essential idea of the excursion set theory; more details 
can be found in Appendix \ref{app:excursion}.

\subsection{Excursion Set Theory}
\label{subsect:excursion}

Assume that the initial local density perturbation filtered at a given scale $R$, $\delta(x, R)$ follows a Gaussian distribution, 
and that there is no correlation of $\delta(x, R)$ between different filter sizes
(for correlated $\delta$, see \citet{ms2012}). Then we know A.) the distribution can be fully described by its variance $S$, and
B.) when varying the filter size $R$ to $R-\de R$ 
or equivalently in hierarchical models, $S\rightarrow S+\de S$, 
the increment of $\delta(x, R)$ is independent from its previous value and should also follow a Gaussian distribution with 
the variance of $\de S$. Thus, $\delta(x, S)$ is just a Brownian motion with `time' variable $S$.
In the spherical collapse model, if a local density exceeds a certain barrier $\delta_c$, then it will collapse and form 
a virialized halo with all the mass $M'$ enclosed within $R'$ by some given time. In the $(S, \delta)$-plane, if we start the walk from the origin, walks that 
cross $\delta_c$ for the first time at $S' = \sigma^2(M')$ correspond to such objects. Walks which cross first at smaller values of $S$ form 
higher mass halos. Therefore, the fraction of mass that has collapsed and formed halos heavier than $M'$ is 
the fraction of random walks $\delta(x,S)$ that have crossed the barrier $\delta_c$ at $S < S'$. Alternatively, 
one can calculate the fraction of mass that is incorporated in halos at a given range of halo mass [$M$, $M+\de M$], 
or equivalently, $[S, S+\de S]$ at a given redshift $z$:

\begin{eqnarray}
f(S,z)\de S = \frac{1}{\sqrt{2\pi S}}\frac{D_+(0)\delta_c}{D_+(z)S}\exp\left[-\frac{D_+^2(0)\delta_c^2}{2D_+^2(z)S}\right]\de S,
\end{eqnarray}
where $f(S,z)$ the first-crossing distribution of the Brownian motion to the barrier $D_+(0)\delta_c/D_+(z)$, and $D_+$ is the linear growth factor.
The first crossing distribution essentially gives the halo mass function (see Appendix \ref{app:excursion}). 
There is equal chance for a random walk to go negative in $\delta$. Thus, once an appropriate first-crossing barrier for voids, $\dv$, is given, one can also find the void size distribution function by the same method.

\subsection{First crossing barrier for void}
\label{subsect:barrier}

For halos, the first crossing barrier $\delta_c$ is usually defined as the linearly extrapolated initial overdensity at 
the time of collapse, i.e., when the mass shells reach zero radius. This time can be calculated using the spherical collapse model. Naively one can find the shell-crossing barrier for 
voids in a similar way. The shell at the radius of $r$ of a perturbed spherical underdense region will expand faster 
than the shell at $r' = r + \Delta r$, as the enclosed mass within the border shell is smaller. 
Shell-crossing for underdense regions occurs when the two shells collide. This occurs at the present day for 
underdense regions with $\dsc=-2.76$ (the density 
contrast at the initial condition extrapolated to today) for the concordance $\rm{\Lambda}CDM$ model. Like $\delta_c$, 
$\dsc$ depends on $\Omega_{\rm m}$ and is independent of smoothing scale. Moreover, the underdense region at shell-crossing 
happens to be very empty, i.e., its nonlinear underdensity is $\delta=-0.8$. Therefore, $\delta=-0.8$ serves nicely as an empirical definition of voids. 

In modified gravity, however, the situation is more complicated. First, 
the shell-crossing barrier can depend on the environment, simply because the fifth force and 
hence the expansion history of shells depends on the environment.
Therefore, one may expect voids (likewise halos \citep{le2012}) to form differently depending on 
the environment.  Second, even for the same environment, the population of voids may also be different from ${\rm \Lambda CDM}$, due to the size dependence of the force which leads to scale dependence of the barrier. 

In chameleon models, the fifth force does speed up the expansion of voids (as seen in Fig.~\ref{fig:single}), but the 
shell-crossing time usually occurs later than in  $\rm{\Lambda}CDM$ with the same initial conditions. This is because 
the effect of the fifth force on the relative accelerations of neighboring shells is in the opposite direction from gravity. For $-1 < \eta < 0$ the fifth force opposes but does not overcome gravity, 
so that the pull of inner shells on outer ones is reduced, making the critical density for shell crossing in chameleon models 
harder to reach. If an observer is riding on the boundary shell, then all the 
nearby shells move closer with time, but more slowly than shells feeling only standard gravity. Furthermore, for some initial 
density perturbations, the shell crossing does not happen at all. 

Since the epoch of shell crossing can be unreasonably late or undefined for these models, it is easier to use empirical criteria for void 
formation. We choose $\delta = -0.8$ as a common criteria for the following reasons, A.) it correspond to the 
first-crossing barrier in ${\rm \Lambda CDM}$, making it easy to compare with results from ${\rm \Lambda CDM}$; 
B.) Voids with $\delta = -0.8$ are indeed very empty, and can be defined by the same way in simulations and observations, thus enabling one to make direct comparisons.
For example, in \citet{Hoyle2004} and \citet{sl2012} they use similar threshold to define voids in the 2dFGRS and SDSS galaxy samples. \citet{pv2012} also find voids from SDSS7 having similar density contrast, $\delta < -0.85$ at the edges. 

\begin{figure*}
\centering
\resizebox{88mm}{!}{\includegraphics{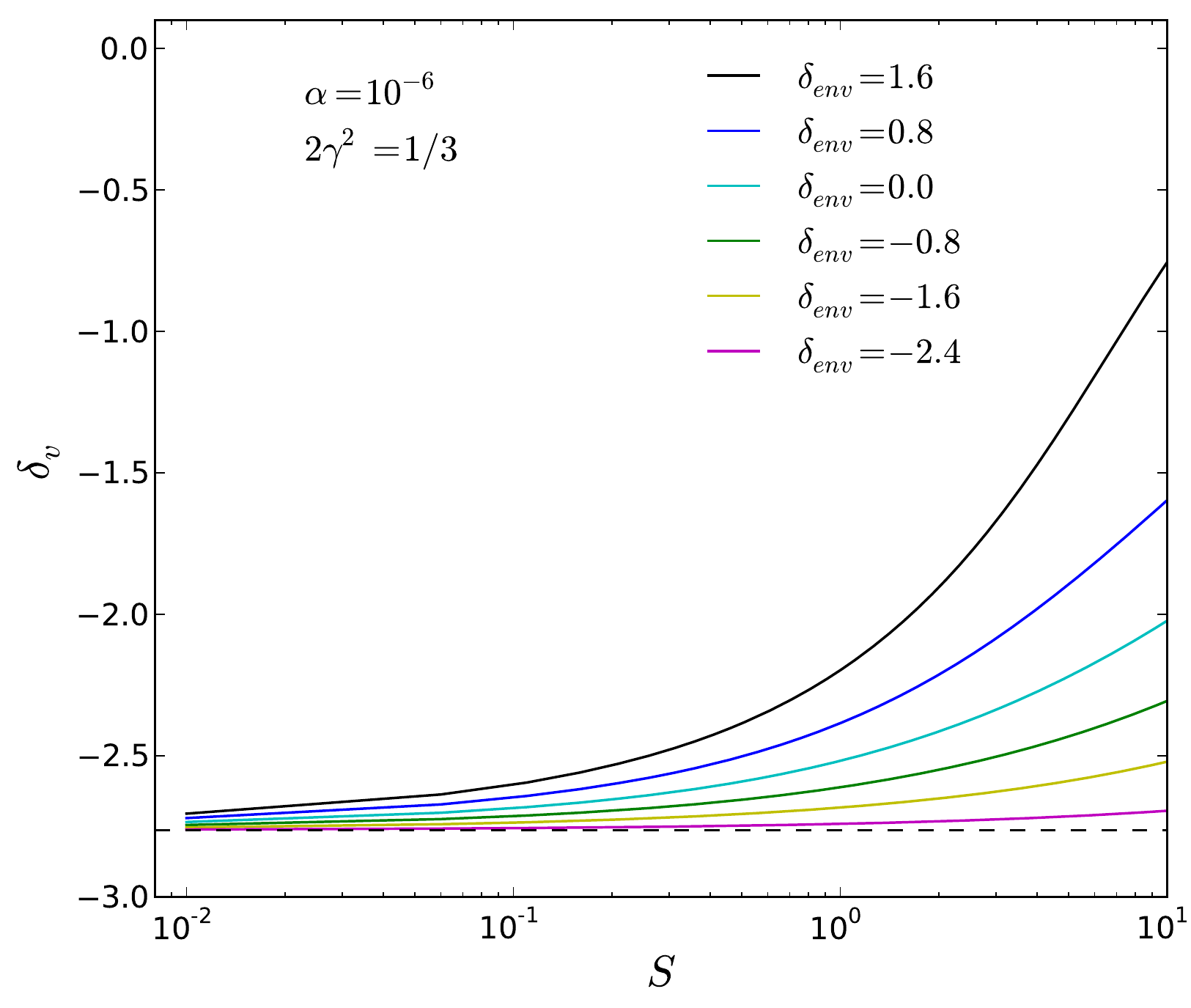}}
\caption{The linearly-extrapolated void formation barriers for various environments as a function of scale, $S = \sigma^2 (M)$. Environment densities decrease from $\denv = 1.6$ to $\denv = -2.4$ from top to bottom. The dashed line shows the constant ${\rm \Lambda CDM}$ barrier, $\delta_{\rm v} = -2.76$ which results from the same void-formation criteria of nonlinear density $\delta = -0.8$.}
\label{fig:barrier}
\end{figure*}

Thus, we use the requirement that the nonlinear density constrast today is $\delta = -0.8$, along with Eqs.~(\ref{eq:yevolution}, \ref{eq:dplus}, \ref{eq:eta-evo}, \ref{eq:yevolution2}), to solve for the initial underdensity as a function of scale $S$ and environment $\denv$. The resulting void-formation barrier is shown in Fig. {\ref{fig:barrier}}.
Unlike ${\rm \Lambda CDM}$ where the crossing barrier is flat, barriers in chameleon models are scale dependent. 
In general, barriers in chameleon models are lower (less negative). Smaller voids have shallower barriers to reach in order 
to form due to the fact that the fifth force in smaller voids is relatively stronger 
(see Fig.~\ref{fig:single}), which makes them to expand faster. In other words, for reaching the same $\delta = -0.8$ today, 
the necessary initial density contrast for smaller voids is smaller (less negative). 
The crossing barriers keep decreasing (becoming less negative) and steepening with the increase of $S$. This is 
very different from the collapsing barrier for halos in the same model, where they are leveling off at $S \sim 5$ \citep{le2012}. 
This difference is a direct result of the fifth force strength upper bound of $2 \gamma^2$, which only applies to overdensities.

Fixing void size, the barrier is lower (less negative) and steeper for denser environment, where the difference 
from the flat barrier in ${\rm \Lambda CDM}$ is also larger. Therefore the difference of void population with ${\rm \Lambda CDM}$ 
should be more prominent in such regions. 
This environmental dependence of crossing barrier is the opposite for halos, where the collapsing barriers are higher 
(more positive), and closer to the ${\rm \Lambda CDM}$ barrier for denser environment \citep{le2012}. 
Qualitatively, these two opposite pictures in voids and halos can be understood by the same reasoning, 
i.e., for voids or halos of the same mass given $\rhoin$ (the mean density in the void or halo region), 
the strength of the fifth force is larger for larger differences between $\rhoin$ and the background density outside 
the perturbed region, $\rhoout$. For voids $\rhoin < \rhoout$, a larger $\rhoout$ means $|\rhoout - \rhoin|$ 
is larger and hence a larger fifth force, while for halos $\rhoin > \rhoout$, a larger $\rhoout$ means $|\rhoout-\rhoin|$ 
is smaller therefore a smaller fifth force.




\subsection{Moving environment approximation}
\label{subsect:moving}

In calculating the void barriers in the previous section, the environment was specified only by its linear density perturbation, $\delta_{\rm env}$. In order to derive the first-crossing distributions and other void statistics it is necessary also to specify an environment length scale.

In treating spherical collapse in chameleon models, \citet{le2012} used an environment scale of 8 Mpc/$h$ for halos of every size. Such a fixed-environment scale works well for halos, since the range of interesting virial radii is fairly small, $\sim 0.1 - 1$ Mpc/$h$. Furthermore, since throughout collapse the proto-halo is always shrinking, there is little worry of its size becoming comparable to the environment scale. On the other hand, the interesting void sizes we are considering range from $\sim 1 - 30$ Mpc/$h$, and each will expand beyond its initial comoving radius by a factor 1.7 at formation. Thus we need to consider the definition of the environment more carefully.  

First, the scale of the environment should at least be larger than the final size of the void. Secondly, it should also be large enough so that the scalar field in the environment has space to settle to its minimum. This is to guarantee that the boundary condition Eq.~(\ref{eq:bc}) for the scalar field profile equation holds. Third, it  cannot be too large because this would simply mean using a value very close to the cosmological density $\bar{\rho}_m$ for all void environments. Bearing these considerations in mind, we introduce a moving environment approximation, in which the initial environment scale is a function of the initial void scale, specifically $\renv = 5 \, R$.


We notice that in the moving environment approximation, the expanding void shell and collapsing environment shell may cross for voids in very overdense environments. 
Therefore we also calculate the first-crossing distributions with a large fixed-environment scale of $\renv = 75 \, {\rm Mpc}/h$, so that the environment shell begins its collapse much farther from the void shell. The difference between the two approximations is less than 10\% for the void scales of observational interest, i.e., those $\sim 1 \, {\rm Mpc}/h$ and larger; details of the comparison can be found in Appendix \ref{app:env}. It follows that the results for choices of $\renv > 5 \, R$ are also less than 10\%, since such environment scales are between our fiducial choice $\renv = 5 \, R$ and the fixed environment scale. This level of difference, as we will see later, is negligible compared to the difference between GR and MG that we are considering. Thus, our main conclusions are insensitive to the definition of environment.

\subsection{Conditional first-crossing distributions}

\subsubsection{Unconditional First Crossing of a Moving Barrier}




The distribution of the first crossing of a general barrier by a
Brownian motion has no analytic solutions except for
some simple barriers, {\it e.g.}, flat \citep{bcek} and linear
\citep{sheth1998,st2002}. Unfortunately  neither of these is
a good approximation to our barriers
in Fig.~\ref{fig:barrier}. As a result, we follow
\cite{zh2006} and numerically compute this distribution. We
briefly review their method for completeness.

Denote the unconditional probability that a Brownian motion starting
off at zero hits the barrier $b(S) > 0$ for the first time in
$[S,S+dS]$ by $f(S)\de S$. Then, $f(S)$, the probability density,
satisfies the following integral equation
\begin{eqnarray} \label{eq:fs}
f(S) &=& g(S) + \int^S_0 \de S'f(S')h(S,S'),
\end{eqnarray}
in which
\begin{eqnarray}
g(S) &\equiv& \left[\frac{b}{S}-2\frac{\de b}{\de S}\right]P\left(b,S\right),\nonumber\\
h(S,S') &\equiv& \left[2\frac{\de b}{\de S}-\frac{b-b'}{S-S'}\right]P(b-b',S-S'),
\end{eqnarray}
where for brevity we have suppressed the $S$-dependence of $b(S)$ and used $b' \equiv b(S')$ and 
\begin{eqnarray}\label{eq:gaussian}
P(\delta,S)\de \delta = \frac{1}{\sqrt{2\pi S}}\exp\left[-\frac{\delta^2}{2S}\right]\de\delta.
\end{eqnarray}
Equation~(\ref{eq:fs}) can be solved numerically on an equally-spaced mesh in $S$: $S_i=i\Delta S$ with $i=0,1,\cdots,N$ and $\Delta S=S/N$. The solution is \citep{zh2006}
\begin{eqnarray}
f_0 &=& g_0 \ =\ 0,\nonumber\\
f_1 &=& (1-\Delta_{1,1})^{-1} g_1,\\
f_{i>1} &=& (1-\Delta_{1,1})^{-1}\left[g_i+\sum_{j=1}^{i-1}f_j(\Delta_{i,j}+\Delta_{i,j+1})\right],\nonumber
\end{eqnarray}
where we have used $f_i=f(S_i)$ and similarly for $g_i$ to lighten the notation, and defined 
\begin{eqnarray}
\Delta_{i,j} &\equiv& \frac{\Delta S}{2}h\left(S_i,S_j-\frac{\Delta S}{2}\right).
\end{eqnarray}
We have checked that our numerical solution matches the
analytic solution for the 
flat-barrier crossing problem.

\subsubsection{Conditional First Crossing of a Moving Barrier}

The unconditional first crossing distribution, which relates directly to the void size distribution function in the $\Lambda$CDM model, is not particularly interesting in the chameleon model. This is because spherical underdensities in different environments will follow different evolution paths. If it is in the environment specified by $(\senv, \denv)$, then $(\senv, \denv)$ should be the starting point of the Brownian motion trajectory. In other words, we actually require the distribution {\it conditional on} the trajectory passing $\denv$ at $S=\senv$;
we write this first-crossing distribution as $f(S, \dv(S, \denv) ~|~ \senv, \denv)$, showing explicitly
the $\delta_{\rm env}$ dependence of $\dv$. The numerical
algorithm to calculate the conditional first crossing probability is a
simple generalization of the one used above to compute the
unconditional first crossing probability \citep{phs2011} and is not
presented in detail here.

Note that the preceding algorithm assumes the barrier $b(S) > 0$, while our void-formation barriers are strictly negative. However, if solving the problem by a Monte Carlo method we could note that the resulting first-crossing distribution is invariant under reflecting the Gaussian random walks about $\delta = 0$ (since each step of each walk is equally likely to move to higher or lower $\delta$). Thus, we can solve the distributions for our negative barriers by using $b(S) = |\dv (S)|$ in the above algorithm.

Furthermore, the preceding algorithm describes the calculation of the first-crossing probability for the fixed-environment approximation, in which a single starting point $(\senv, \denv)$ for a given barrier $\dv(S, \denv)$ is sufficient. To implement the moving environment approximation we calculate a new first-crossing distribution for each underdensity scale $S$, where the walk starts at $\senv (\renv)$ and $\renv = 5 R(S)$ as described in Sec.~\ref{subsect:moving}.
Our final result for the conditional first-crossing probability is then $f(S, \dv(S, \denv) ~|~ \senv(S), \denv)$, where the dependence of $\senv$ on $S$ is written explicitly. 


In the special case where the barrier is flat, $\delta_{v}(S, \delta_{\rm env})=\dsc$, $f(S, \dv(S,\delta_{\rm env}) ~|~ \senv(S), \denv)$ is known analytically as
\begin{eqnarray}
f &=& \frac{|\dsc-\delta_{\rm env}|}{\sqrt{2\pi}\left(S-\senv \right)^{3/2}}\exp\left[-\frac{\left(\dsc-\delta_{\rm env}\right)^2}{2\left(S-\senv \right)}\right],
\end{eqnarray}
where again $\senv = \senv(S)$, so that in the next section we compare first-crossing distributions for GR and MG both calculated using the same moving environment scale.

\subsubsection{Results}

\begin{figure*}
\centering
\resizebox{180mm}{!}{\includegraphics{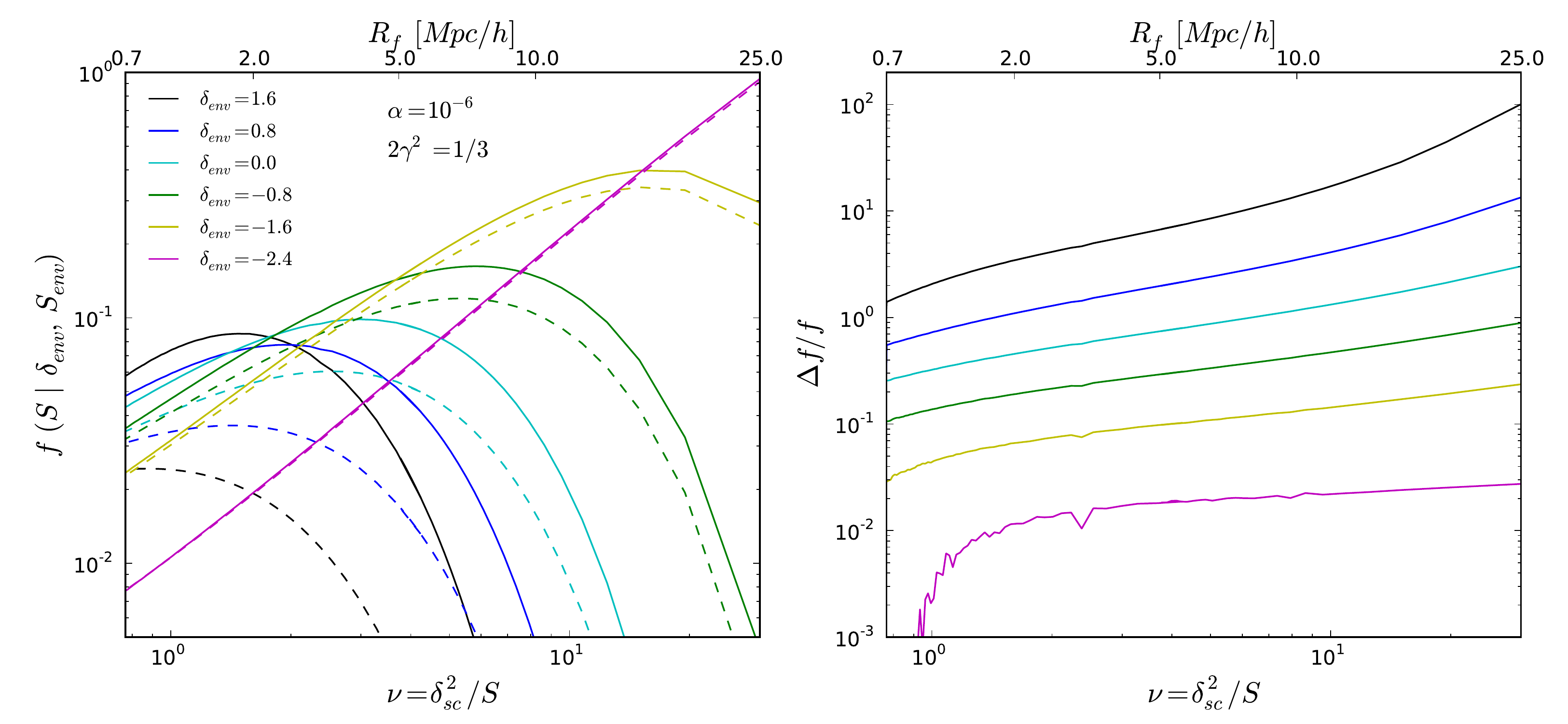}}
\caption{{\it Left:} First crossing distribution functions for different environments as indicated by $\denv$. 
Solid lines are in chameleon cosmology with our fiducial model parameters.  The top $x$-axis labels the corresponding final 
void radius when the density contrast of a void reaches $\delta=-0.8$. In bottom $x$-axis, $\dsc = -2.76$ is the 
shell-crossing barrier for voids in $\Lambda$CDM, and $S=\sigma^2(M)$ is the variance of a spherical top-hat region. 
{\it Right:} Fractional differences of the first crossing distributions 
between GR and chameleon cosmology for different environments.}
\label{fig:f_cond}
\end{figure*}

Figure \ref{fig:f_cond} shows the first-crossing distribution of voids in different environments. In general, we find 
all voids today with radii $R_f \gtrsim 1$ Mpc$/h$ are more numerous in chameleon models, for all environments. This difference from 
${\rm \Lambda CDM}$ is 
larger for overdense environments. 
This is a consequence of previous results of this paper, namely that the fifth force is 
relatively stronger for denser environments. 

Next, consider fixing the environment density. In this case, the fractional difference 
of the number density between chameleon models and GR tends to be greater for larger voids (larger $\nu \equiv \dsc^2 / S$ or smaller $S$), as indicated by 
the increase of $\Delta f/f$ with $\nu$ in the figure. For example, in the environment of $\delta_{\rm env}=0.8$, voids with $R_f=5$ Mpc$/h$ 
may be 2 to 3 times more common than those in ${\rm \Lambda CDM}$, and 10 times more for $R_f=25$ Mpc/$h$. This difference may seem surprisingly large, but such a case may be too rare to be observed. If one smooths the initial density field with a filter size much 
greater than $R = 15$ Mpc/$h$ (corresponding to $R_f=25$ Mpc/$h$), the probability distribution of the overdensity will be a narrow Gaussian with zero mean. The chance of 
having a linearly-extrapolated $\denv=0.8$ should be very low; the odds of such an environment developing voids of $R_f=25$ Mpc/$h$ or larger with $\delta=-0.8$ will be even less. Therefore, it might be difficult to find large voids in very overdense environments, where the predicted difference between models is expected to be larger. In reality, most large-scale environments are very close to the cosmic mean, i.e., $\denv \sim 0$. In this case, 
the difference between models indicated by $\Delta f/f$ is less extreme but still very significant, being $\approx 100\%$ for $R_f=5$ Mpc/$h$ and $\approx 300\%$ for $R_f=25$ Mpc/$h$. We shall see in the next subsection that this difference is indeed close to the case where the average over all environments is taken.

The environmental dependence of model differences in the conditional first crossing distribution of voids is just the opposite as that for halos for 
reasons we have explained in Sec.~\ref{subsect:barrier}. The halo mass function \citep{le2012} is found to differ more from its ${\rm \Lambda CDM}$ 
counterpart in underdense environments. 

The fact that $\Delta f/f$ is larger for larger voids might seem counter-intuitive, as we have shown that 
the relative strength of the fifth force is smaller for larger voids, hence the difference in their expansion velocities and sizes today are relatively smaller. However, the difference in the number density of voids is also related 
to the shape of the void size distribution function.
Consider that $f$ is a very steep function of $\nu$ when $\nu$ is large. A small increment 
in $R_f$ or $\nu$ can therefore lead to a relatively large change in $f$. 

In principle, if $f_{\rm MG}$ is larger than $f_{\rm GR}$ for large voids, the opposite should be true for small voids, namely the abundance 
of small voids will be lower in chameleon models. This is expected from the normalization of the first-crossing probability. Picturing 
this in the excursion set theory, in chameleon theories Brownian motions are likely to cross the barrier at a 
slightly earlier `time', i.e. small $S$, corresponding to large voids. Correspondingly, the probability of a Brownian motion 
to survive for longer and cross the barrier at large $S$ is reduced -- voids of smaller sizes are (relatively) rarer than in 
${\rm \Lambda CDM}$. Therefore, the solid and dashed lines in Fig.~\ref{fig:f_cond} will cross each other at some small $\nu$ that is not plotted, 
namely the abundance of small voids can be lower in chameleon models. In fact, such a crossing point 
is also expected for halos, which has been shown to be at $S \lesssim 10$ for 
the environments under consideration \citep{le2012}. For voids, the crossing points are found to appear 
at much larger $S$. This is likely due to the halo barriers leveling off at $S \sim 5$, while the void barriers continue 
to steepen towards larger $S$. 

In real observations, one needs to have tracers like galaxies or galaxy clusters to define void walls. If the size of the void is 
comparable to that of the tracers, then the walls will be lumpy. 
Voids with radii comparable or smaller than the typical size of virialized objects are therefore not well defined and of little interest. 
We do not show results deeply into this regime. In the range of empirical interest, we only see the lines of $f_{\rm MG}$ and $f_{\rm GR}$ crossing each other 
for the case of $\denv = -2.4$ at $R \sim$ 1 Mpc/$h$, which should be a rare situation. Thus, for denser 
environments we always expect to find more voids in chameleon models at all empirically meaningful sizes.  

The environmental dependence of the differences between models may provide useful guidelines for testing gravity. In overdense environments, one 
may want to look at the statistics of large voids as the difference with ${\rm \Lambda CDM}$ may be larger, 
while in underdense regions, the difference in halo population may be larger therefore halo number densities may be more interesting to analyze. 
We summarize these two cases as void-in-cloud and cloud-in-void. However, both of these two cases are relatively uncommon to find in the real 
universe so that the statistics may be poor. In this case, using most of the observed volume could provide better constraints since the sample of voids and 
halos would be larger. It is therefore interesting to determine the overall difference between models once we average over 
all different environments.

\subsection{Environment-averaged first-crossing}



To get the average first crossing distribution of the moving barrier, we must integrate over all environments. The distribution of $\delta_{\rm env}$, denoted as $q(\delta_{\rm env}, \delta_{\rm c}, \senv)$, in which $\delta_{\rm c}$ is the critical overdensity for spherical collapse in the $\Lambda$CDM model, is simply the probability that the Brownian motion passes $\delta_{\rm env}$ at $\senv$ and never exceeds $\delta_{\rm c}$ for $S<\senv$ (because otherwise the environment itself has collapsed already). This has been derived by \cite{bcek}:
\begin{eqnarray}\label{eq:q}
q(\delta_{\rm env},\delta_{\rm c},\senv) &=& \frac{1}{\sqrt{2\pi \senv}}\exp\left[-\frac{\delta^2_{\rm env}}{2\senv}\right]\nonumber\\
&&-\frac{1}{\sqrt{2\pi \senv}}\exp\left[-\frac{\left(\delta_{\rm env}-2\delta_{\rm c}\right)^2}{2\senv}\right],
\end{eqnarray}
for $\delta_{\rm env} \leq \delta_{\rm c}$ and $0$ otherwise. Again, we have $\senv = \senv(S)$, so that the distribution $q$ changes for each void size. For smaller smoothing length (larger $\senv$), the pdf of $\denv$ is wider so that the very overdense and very underdense environments are more likely to be sampled.

Then the environment-averaged first crossing distribution will be
\begin{eqnarray}\label{eq:f_ave}
f_{\rm avg}(S) &=& \int^{\delta_{\rm c}}_{-\infty}q\times f(S, \dv(S,\denv) ~|~ \senv(S), \denv) \, \de \denv.
\end{eqnarray}
The environment-averaged first-crossing distribution and void volume function are related by
\be
\frac{\de n}{\de V} \de V = \frac{\bar{\rho}_m}{M} f_{\rm avg} (S) \left| \frac{\de S}{\de V}\right| \de V \, ,
\ee
where V is the final volume of the void given by
\be
V = \frac{M}{\bar{\rho}_m} \times (1.71)^3 \, .
\ee
The factor of $1.71^3$ results from our void formation criteria of nonlinear density $\delta = -0.8$. By mass conservation, such an underdensity which was originally at the cosmic mean has grown to 1.71 times its initial comoving radius.

\begin{figure*}
\centering
\resizebox{170mm}{!}{\includegraphics{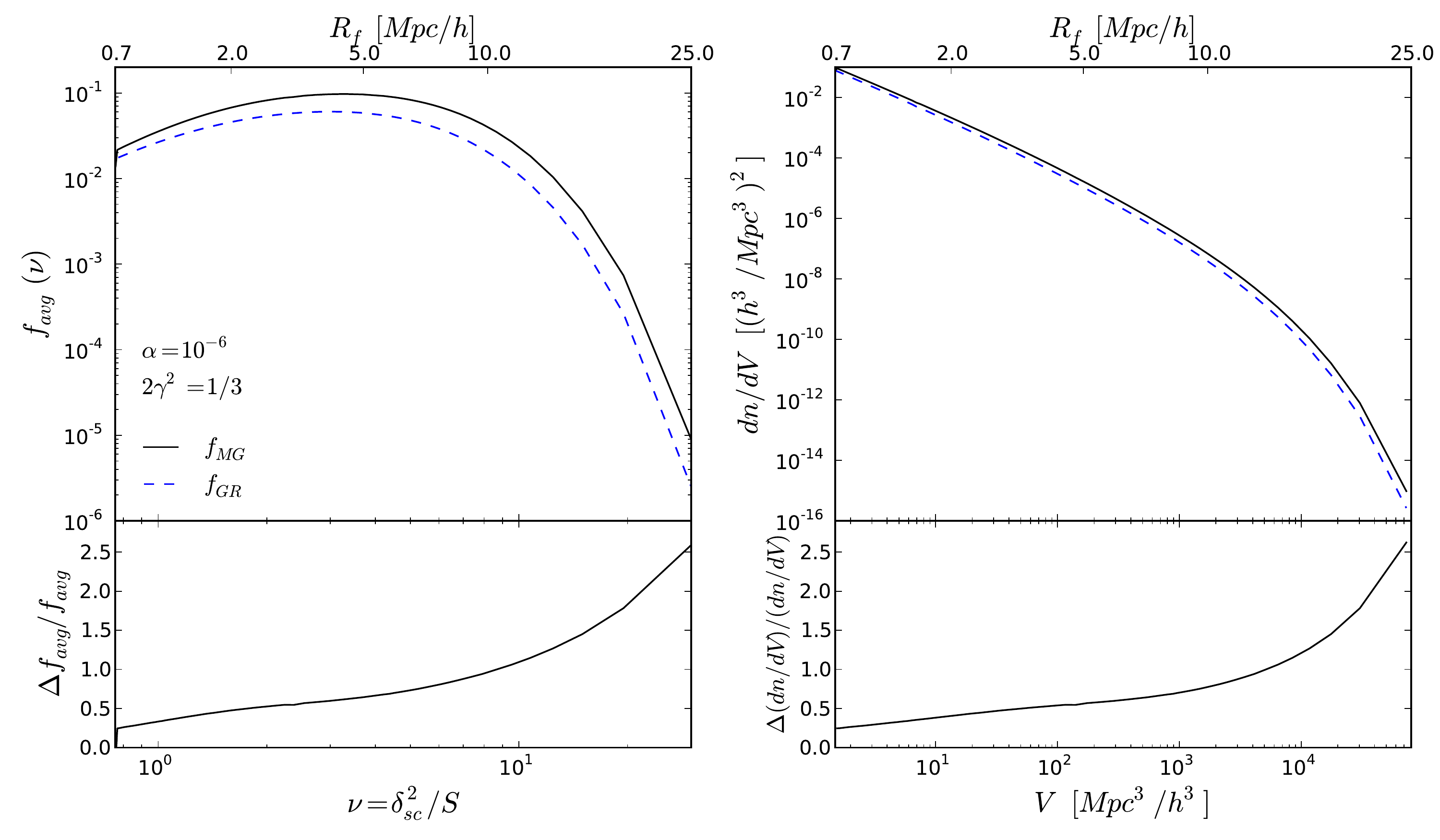}}
\caption{{\it Left panel:} Compares the averaged first crossing distribution functions between chameleon model (solid line) 
and GR (dashed line). The fractional difference is shown in the bottom panel.
{\it Right panel:} Void volume distribution functions and their fractional difference. The difference in the number density of voids between the two models increases monotonically with void size.}
\label{fig:f_avg}
\end{figure*}

The left and right panels of Fig.~\ref{fig:f_avg} show the environment-averaged first-crossing distribution and the corresponding void volume function, respectively. Comparing the environment-averaged void distribution functions between our fiducial chameleon model and $\Lambda$CDM, 
we find the fractional difference in the number density of voids between the two models increases with void size. At $R_{f}\sim 25$, one may expect to find 2 
to 3 times more voids in chameleon models, and such a difference will keep increasing for larger voids. This level of difference in the void population is much 
greater than that in halos, where the difference of mass function is found to be no more than $20\%$ \citep{le2012}: a factor of 10 times smaller difference.
The boost of probability for having large voids in chameleon models has interesting implications for observation, thus serving as a powerful test of gravity theories. 
Given a finite survey volume, one can simply count the number of voids greater than a certain radius, e.g., $R_f > 25$ Mpc/$h$ 
to find out the number density of them and then compare it with different models.

\subsection{Theory Variations}
\label{sect:vary-ag}

\begin{figure*}
\centering
\resizebox{180mm}{!}{\includegraphics{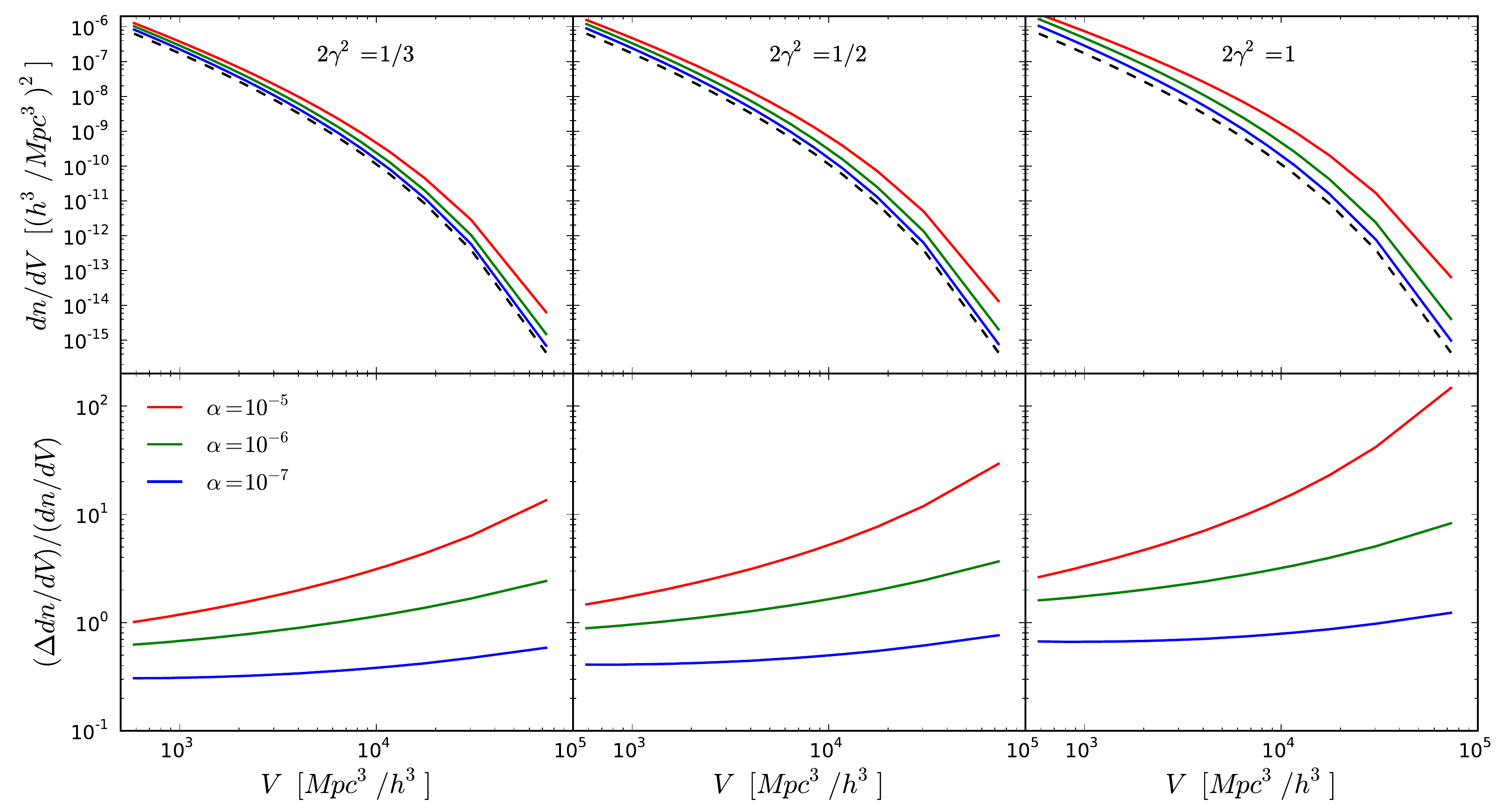}}
\caption{{\it Left panels}: The void differential volume function for our chameleon model with $2\gamma^2 = 1/3$ (solid lines) compared to GR (dashed). Various values of $\alpha$ are shown, ranging over $10^{-5}, 10^{-6}$, and $10^{-7}$, from top to bottom. The lower panel shows the fractional difference from the GR result.
{\it Center panels}: The same, but for coupling $2\gamma^2 = 1/2$.
{\it Right panels}: The same, but for coupling $2\gamma^2 = 1$. Even for $\alpha = 10^{-7}$, $2\gamma^2 = 1/3$ where the deviation is weakest, it is above 30\% for all empirically interesting void sizes.}
\label{fig:dndv-vary}
\end{figure*}

Up to this point, we have only shown results for our fiducial chameleon theory, with parameters $\alpha = 10^{-6}$ and $2 \gamma^2 = 1/3$. Figure~\ref{fig:dndv-vary} shows the effect of varying these two parameters on the volume function, $\de n / \de V$. Focusing on the leftmost panels we see the models with $2\gamma^2 = 1/3$, which correspond most closely to the $f(R)$ class of theories. In moving from $\alpha = 10^{-5}$ to $10^{-7}$ the fractional difference changes by a factor $\sim 3$ for small voids ($V = 7\times 10^{2}$ (Mpc$/h)^3$) and by $\sim 25$ for voids two orders of magnitude larger ($V = 7\times 10^{4}$ (Mpc$/h)^3$).

A direct comparison of this chameleon theory with $f(R)$ models is not possible, but comparing the compton wavelengths can give some idea of the differences. For the $f(R)$ model of \citet{hs2007}, the compton wavelength in the background density today is $\sim 3$ Mpc$/h$ for $|f_{R0}| \sim 10^{-6}$. Our fiducial model has a longer compton wavelength: for $2\gamma^2 = 1/3$ we have $\lambda \sim 2 \sqrt{10^8 \alpha}$. Thus for $\alpha = 10^{-6}$, $\lambda \sim 20$ Mpc$/h$.

As it is also interesting to put constraints on the coupling $2\gamma^2$ (e.g., \citep{jv2012}), we show such variations in the center and right panels of Fig.~\ref{fig:dndv-vary}. As we expect, for fixed $\alpha$ the deviations are much larger for stronger couplings. Again the largest, rarest voids are most sensitive to these changes due to the steepness of the volume function there: for $\alpha = 10^{-5}$ the deviation of the volume function from the GR result grows by a factor of 10 in moving from $2\gamma^2 = 1/3$ to $2\gamma^2 = 1$.

The comparison to results for the excursion-set mass function highlights the promise of using voids to constrain modified gravity. Consider the case in Fig.~\ref{fig:dndv-vary} with the smallest deviations from GR, $\alpha = 10^{-7}$ and $2\gamma^2 = 1/3$. The fractional difference in the volume function is 30-60\% over at least two decades in void volume. The deviation of the mass function predicted by this model peaks at 5\% for halo masses $\sim 10^{13} \msun/h$, falling quickly for smaller and larger halos \citep{le2012}. Thus, if the difference between models is integrated over the entire range of halo and void number densities, the total constraining power of the void statistics will be much greater. This larger difference in the void statistics is a result of several effects: A.) the upper bound in the ratio of gravity and the fifth force does not apply to underdensities and B.) the crossing point of GR and MG first-crossing distributions expected due to the normalization of the distribution occurs for voids which are too small to be empirically relevant. Thus, the MG void volume function shows large deviations at all void sizes.

In Appendix~\ref{app:vary-ag} we discuss the effect of varying $\alpha$ and $\gamma$ on the conditional first-crossing distributions, i.e., before the environment averaging is carried out.

\section{conclusion and discussion}
\label{sect:conclusion}
We have explored the physics of the fifth force in voids for chameleon models and applied it to understand 
the impact on void properties. In scalar-tensor theories, such as chameleon MG, the smooth part of the scalar field is the source of the cosmological
constant, known to act like a repulsive force. This is common in both a $\Lambda$CDM universe and a chameleon 
universe. The coupling of the scalar field to mass density causes an additional spatial 
fluctuation of gravity, i.e., the fifth force. This is the only difference for void evolution between chameleon and $\Lambda$CDM models. 
The evolution of voids in MG is affected by the Newtonian force, the dark energy force and the fifth force. 

The following interesting features are found in comparison to 
a $\Lambda$CDM universe, some of which may be used to test gravity in laboratory experiments and observational data, or to guide more precise predictions from cosmological $N$-body simulations.
\\
\\
1.) The fifth force in voids is a type of `anti-gravity'. It points outwards from the center of the void, opposite to the 
direction of normal gravity. This is because the slope of the scalar field profile is negative in voids.
\\
\\
2.) In principle, the amplitude of the fifth force 
can be very large in voids. In halos, its magnitude can be no more than $2\gamma^2$ (1/3 in our fiducial model) of normal gravity. Due to the breakdown of Birkhoff's theorem, the scalar field profile and hence the fifth force are functions of the matter density inside and outside the void region, as well as its size. 

This property leads us to suggest a possible laboratory test of gravity 
using a vacuum chamber. To create a chamber inside of which the fifth force is dominant, it should have a thick chamber wall made of 
high density material. This is to enlarge the density contrast between the chamber interior and the wall so that fifth force strength is maximized. 
The wall needs to be thick to have enough space for the scalar field to reach its minimum in the wall.
Walls of the chamber and test particles (detectors) in the chamber should feel the fifth force pushing outwards, but very little gravity (as long as the chamber is close to real vacuum).
In the neighborhood of the earth, the background density is non-zero. There is dark matter from the Milky 
Way halo, and maybe some baryonic dust; these two should contribute a haze of mass density inside any vacuum 
chamber. This may set the limit for the density contrast and the amplitude of the fifth force. Furthermore, although the ratio $F_5 / F_{\rm N}$ may be large in this case, we know that $F_{\rm N}$ is quite small in the chamber, so that the large ratio does not necessarily imply a large fifth force. The effect of the dark energy force also needs to be accounted for. We leave the quantitative investigation of this experiment to future work.
\\
\\
3.) Driven by the additional fifth force, individual voids expand faster and grow larger than their ${\rm \Lambda}$CDM 
counterparts. The fractional difference in void radius and expansion velocity is larger for small voids in overdense 
environment (void-in-cloud), at the level of $\lesssim 10\%$ and $20 - 30\%$ respectively, for voids of a few Mpc/$h$. For the same reason, 
voids of the same size should be emptier in chameleon models. This leads to interesting observational consequences. A.) In redshift space, 
due to the faster expansion of voids, a small void-in-cloud may be more elongated along the line-of-sight due to redshift space distortion. 
B.) Void profiles may be steeper as voids empty themselves more quickly, as has been shown in \citet{ms2009}.
We plan to investigate both of the above by stacking voids in simulations. Recent work has shown that the lensing signal from stacked voids in future surveys will provide information on their radial profile \citep{kc2012}. This may provide a complimentary probe to void statistics for distinguishing between gravity models.
\\
\\
4.) For individual voids, the largest difference between GR and MG is found in void-in-cloud systems, while for voids statistics, the large voids differ more. 
The fractional difference in the number density of voids increases with size and is $\sim$10 times larger than the corresponding difference for halos.
The chance of having voids with $\delta\sim -0.8$ with $R\sim 25$ Mpc/$h$ is 2.5 times larger than in ${\rm \Lambda}$CDM. A conceptually simple observational test would be to count the number of very large halos in a volume limited sample, and find out the probability for that count to occur within different gravity models. 

In fact, the detection of the CMB Cold Spot in the WMAP data, if interpreted as the ISW signal, has already imposed a constraint on this probability. 
The size of the void in the large-scale structure needed to generate the size and amplitude of the Cold Spot is estimated to be at the order of
$100$ Mpc/$h$ in radius, which may not be consistent with a ${\rm \Lambda}$CDM universe \citep[e.g.][]{
Cruz05, Inoue06, Rudnick07, Masina09, Cai2010}. Similarly, the detected ISW signal from the 
stacking of 4-deg$^2$-size regions of the CMB corresponding to the SDSS super clusters and super voids is found to be $2-3 \sigma$ higher than 
that expected in a ${\rm \Lambda}$CDM universe \citep{Granett08, Papai11}. Recent work has shown that the abundance of the largest voids in $\Lambda$CDM simulations 
may be too small to match observations \citep{tv2012}. All of these discrepancies, if confirmed, seem to indicate that very large structures in the 
universe are perhaps larger and more abundant than expected in a ${\rm \Lambda}$CDM universe. The fact  that the abundance of large voids in modify 
gravity can be much greater than in ${\rm \Lambda}$CDM suggest that modify gravity can somewhat release the tension imposed by those observations, 
but precise quantitative predictions are beyond the reach of the spherical collapse model and excursion set theory.
 
Intriguingly, there are also observations suggesting that 
galaxies are less common in low density regions than expected in the standard cosmology \citep[e.g.][]{Tikhonov09}. The Local Void (within the radius of 1-8 Mpc from the center of the local group) also seems far too empty based on the galaxy number density \citep[e.g.][]{Peebles2001, Tikhonov09, Peebles2010}, but see \citet{Tinker09} for a different view. 
There is also an unexpected presence of large galaxies on the outskirts of the Local Void \citep{Peebles2010}. ``These problems would be eased if structure grew more rapidly than in
the standard theory, more completely emptying the Local Void and piling up matter on its outskirts'' \citep{Peebles2010}. 
Voids in chameleon models seem to coincide qualitatively with these observations. However, the complexity of galaxy formation, 
especially its dependence on environment, is a hard barrier to overcome before any conclusive results can be drawn.
\\
\\
We note that our results for void statistics should be qualitatively similar in other models with chameleon screening, such as $f(R)$ \citep{hs2007}. Furthermore, while symmetron \citep{hk2010, hk2011} and environmentally-dependent dilaton \citep{dp1994, bbds2010} theories rely on conceptually different mechanisms to screen the fifth force, the qualitative picture of Fig. \ref{fig:scalar} is unchanged. The minimum of the symmetron and dilaton fields will again be higher inside an underdensity than outside, thus leading to a repulsive fifth force which will aid the dark energy in speeding up void growth.
\\
\\
\textbf{Caveats:}
\\
\\
Throughout the paper, we employ the spherical collapse model and excursion set theory for studying the evolution of individual voids 
and their distribution functions. However:

1. Voids in the real universe are not perfectly spherical. \citep[e.g.][]{Shandarin2006}.

2. The excursion set theory for voids may not be able to match precisely voids found from simulation or observation. 
There are obvious reasons for this:. A.) It has been noticed that the total volume of voids given by this model 
exceeds that of the universe \citep{sv2004}. This is certainly not physical. One obvious reason is that some `voids' may be embedded 
in overdense regions  whose density reaches the collapsing barrier. This is the void-in-cloud problem, which is more acute for 
small voids. Accounting for it can resolved the problem to some extent, but not fully \citep{sv2004, pls2012}. Another reason is that there is an 
underlying assumption that voids can expand forever, which is also unphysical. The expanding walls of voids will certainly meet their 
neighbors and cross each other. This is probably more complicated to fix and is beyond the scope of this paper.

3. Our results are for voids in the dark matter distribution, whereas observed voids must be defined with respect to galaxies. The excursion set theory of the void population has been extended to these more empirically relevant voids by \citet{fp2006}.

In this paper, we are mostly comparing the difference between two models rather than the accuracy of each model itself. Thus, these well-known limitations of the basic excursion set theory of voids may affect MG and $\Lambda$CDM in roughly the same way, leaving the difference mostly unaffected.
We therefore neglect these problems, and leave the calibration of the theory to simulation for future work. 


\section*{Acknowledgments}

We would like to thank Bhuv Jain, Justin Khoury, Tsz Yan Lam, and Adam Lidz for helpful discussions and comments.
YC acknowledge the support from the NASA grant NNX11AI25G and the Durham International Junior Research Fellowship.

\appendix

\section{Excursion Set Theory}
\label{app:excursion}

It is widely accepted that the large-scale structure (LSS) in the
Universe has developed hierarchically through gravitational
instability.  The excursion sets (regions where the matter
density exceeds some threshold when filtered on a suitable scale)
generally correspond to sites of formation of virialised structures
\citep{ss1988,ck1988,ck1989,efwd1988,er1988,nw1987,cc1988}.

The filtered, or smoothed, matter density perturbation field
$\delta({\bf x}, R)$, is given by
\begin{eqnarray}\label{eq:smooth}
\delta({\bf x}, R) &=& \int W(|{\bf x}-{\bf y}|; R)\delta({\bf y})\de^3{\bf y}, \nonumber\\
&=& \int \tilde{W}(k; R)\delta_{\bf k}e^{i\bf{k\cdot x}}\de^3{\bf k},
\end{eqnarray}
where  $W({\bf r}; R)$ is a filter, or window function,
 with radius $R$, and $\tilde{W}(k;R)$ its Fourier transform; $\delta({\bf x})\equiv\rho({\bf x})/\bar{\rho}-1$ is the 
true, unsmoothed, density perturbation field and $\delta_{\bf k}$ its Fourier transform; we will always use an overbar to denote background quantities.

As usual, we assume that the initial density perturbation field
$\delta({\bf x})$ is Gaussian and specified by its power spectrum
$P(k)$. The root-mean-squared (rms) fluctuation of mass in the
smoothing window is given by
\begin{eqnarray}\label{eq:rms_fluc}
S(R)\ \equiv\ \sigma^2(R)\ \equiv\ \langle\delta^2({\bf x}; R)\rangle\ =\ \int P(k)\tilde{W}(k; R)\de^3{\bf k}.
\end{eqnarray}
 Note that, given the power spectrum $P(k)$, $S$, $R$ and $M$ are
 equivalent measures of the scale of a spherical perturbation and they
 will be used interchangeablly below.

If $\tilde{W}(k;R)$ is chosen to be a sharp filter in k-space, then
the increment of $\delta({\bf x};R)$ as $R\rightarrow R-\delta R$
or equivalently $S\rightarrow S+\delta S$ comes from only the extra
higher-$k$ modes of the density perturbation (see
Eq.~(\ref{eq:smooth})). The absence of correlation between these
different wavenumbers 
means that the increment of $\delta({\bf x};R)$ is independent of its
previous value. It is also a Gaussian field,
with zero mean and variance $\delta S$. Thus, considering $S$ as a
`time' variable, we find that $\delta({\bf x};S)$ can be described by
a Brownian motion.

The probability distribution of $\delta({\bf x};R)$ is a Gaussian
\begin{eqnarray}\label{eq:gaussianApp}
P(\delta,S)\de\delta = \frac{1}{\sqrt{2\pi S}}\exp\left[-\frac{\delta^2}{2S}\right]\de\delta.
\end{eqnarray}
In an Einstein-de Sitter or a $\Lambda$CDM universe, the linear growth
of initial density perturbations is scale-independent, so that
$\delta({\bf x})$ and $\sigma(R)=\sqrt{S}$ grow in the same manner,
and as a result the density field will remain Gaussian while
it is linear. Following the
standard literature, hereafter we shall use $\delta({\bf x}; R)$ to
denote the initial smoothed density perturbation extrapolated to the
present time using linear perturbation theory, and the same for
$\sigma$ or $S$.

In the standard cold dark matter scenario, the initial smoothed
densities which, extrapolated to the present time, equal (exceed)
$\delta_c$ correspond to regions where virialised dark matter halos
have formed today (earlier). In an Einstein-de Sitter universe
$\delta_c$ is a constant, while in a $\Lambda$CDM universe it depends
on the matter density $\Omega_m$. In neither case does $\delta_c$
depend on the size of (or equivalently the mass enclosed in) the
smoothed overdensity, or the environment surrounding the overdensity.

As a result, to see if a spherical region with initial radius $R$ has
collapsed to virialised objects today or lives in some larger region
which has collapsed earlier, we only need to see whether $\delta({\bf
  x}; \geq R)\geq\delta_c$. Put another way, the fraction of the
total mass that is incorporated in virialised dark matter halos
heavier than $M=\frac{4}{3}\pi R^3\bar{\rho}_i$ is just the fraction
of the Brownian motion trajectories $\delta({\bf x}; S)$ which have
crossed the constant barrier $\delta_c$ by the `time' $S=S(R)$, which
is given by \citet{bcek}
\begin{eqnarray}
F(M,z) = \frac{1}{\sqrt{2\pi S}}\int^\infty_{\frac{D_+(0)}{D_+(z)}\delta_c}\left[e^{-\frac{\delta^2}{2S}}-e^{-\frac{(\delta-2\delta_c)^2}{2S}}\right]\de\delta,
\end{eqnarray} 
where the lower limit of the integral is $\frac{D_+(0)}{D_+(z)}\delta_c$, because if a virialised object formed at redshift $z$, then its corresponding initial smoothed density linearly extrapolated to $z$ is 
$\delta_c$, while extrapolated to today it is $\frac{D_+(0)}{D_+(z)}\delta_c$ with $D_+(z)$ being the linear growth factor at $z$. In Einstein-de Sitter cosmology $D_+(z)\propto(1+z)^{-1}$ and
this quantity becomes $(1+z)\delta_c$.

Alternatively, one can say that the fraction of the total mass that is
incorporated in halos,  the radii of which fall in $[R,R+\delta R]$ (or
equally $[S,S+\delta S]$) and which collapse at $z=z_f$ is given
by
\begin{eqnarray}
f(S,z_f)\de S = \frac{1}{\sqrt{2\pi S}}\frac{D_+(0)\delta_c}{D_+(z_f)S}\exp\left[-\frac{D_+^2(0)\delta_c^2}{2D_+^2(z_f)S}\right]\de S,
\end{eqnarray}
where $f(S)$ the distribution of the first-crossing time of the Brownian motion to the barrier $D_+(z=0)\delta_c/D_+(z=z_f)$. Once this is obtained, one can compute the halo mass function observed at $z_f$ as
\begin{eqnarray}
\frac{\de n(M)}{\de M}\de M = \frac{\bar{\rho}_m(z_f)}{M}f(S)\de S.
\end{eqnarray}
Other observables, such as the dark matter halo bias \citep{mw1996} or
merger history \citep{lc1993}, can
be computed with certain straightforward generalizations of the theory.

\section{Comparing fixed- and moving-environment models}
\label{app:env}

\begin{figure*}
\centering
\resizebox{80mm}{!}{\includegraphics{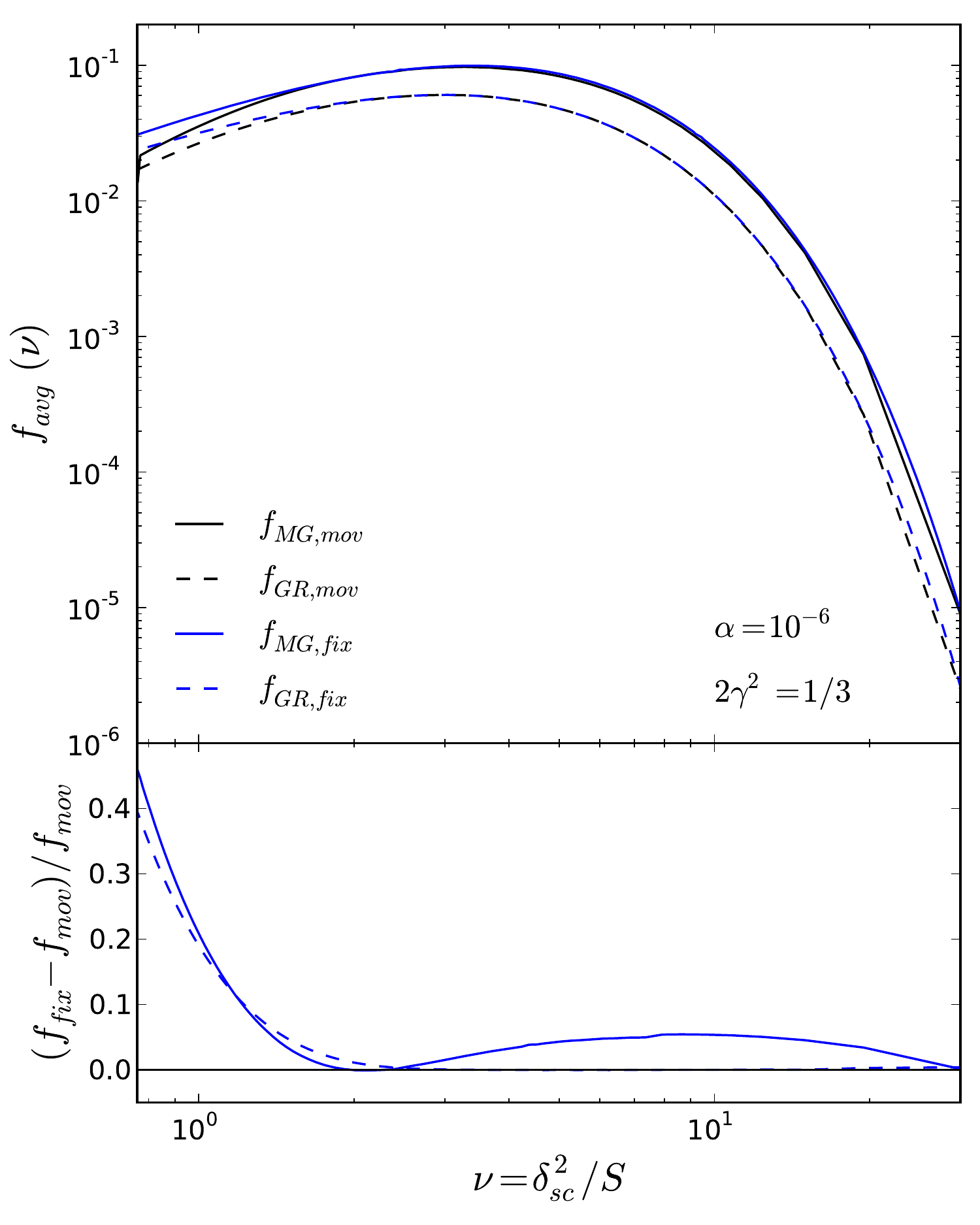}}
\caption{{\it Upper panel:} Environment-averaged first-crossing distribution of voids with (solid lines) and without (dashed lines) the fifth force. The higher solid and higher dashed lines show results for the fixed-environment approximation, while the lower pair show the moving environment approximation.
{\it Lower panel:} Fractional difference of fixed- from moving-environment approximation for modified gravity (solid) and GR (dashed). For $\nu \sim 1$ and larger, the observational range of interest, the difference is below 10\%.}
\label{fig:fixed}
\end{figure*}

In the main text we use a moving-environment approximation in which the smoothing scale of the environment is a function of the void scale, specifically $\renv = 5 R$. However we have also checked the effect of using a fixed-environment approximation to calculate the fifth force. We compare the effect of the approximations on the environment-averaged first-crossing distribution in Fig.~\ref{fig:fixed}, for a fixed-environment scale of $\renv = 75 \, {\rm Mpc}/h$, corresponding to $\senv = 0.01$. The differences are below 10\% for $\nu \gtrsim 1$, corresponding to final void radii $R_f \gtrsim 1 \, {\rm Mpc}/h$. Thus, throughout the range of observable void sizes our conclusions are fairly insensitive to the precise approximation used to calculate the environmental effect of the fifth force.

\section{Theory variations}
\label{app:vary-ag}

\begin{figure*}
\centering
\resizebox{180mm}{!}{\includegraphics{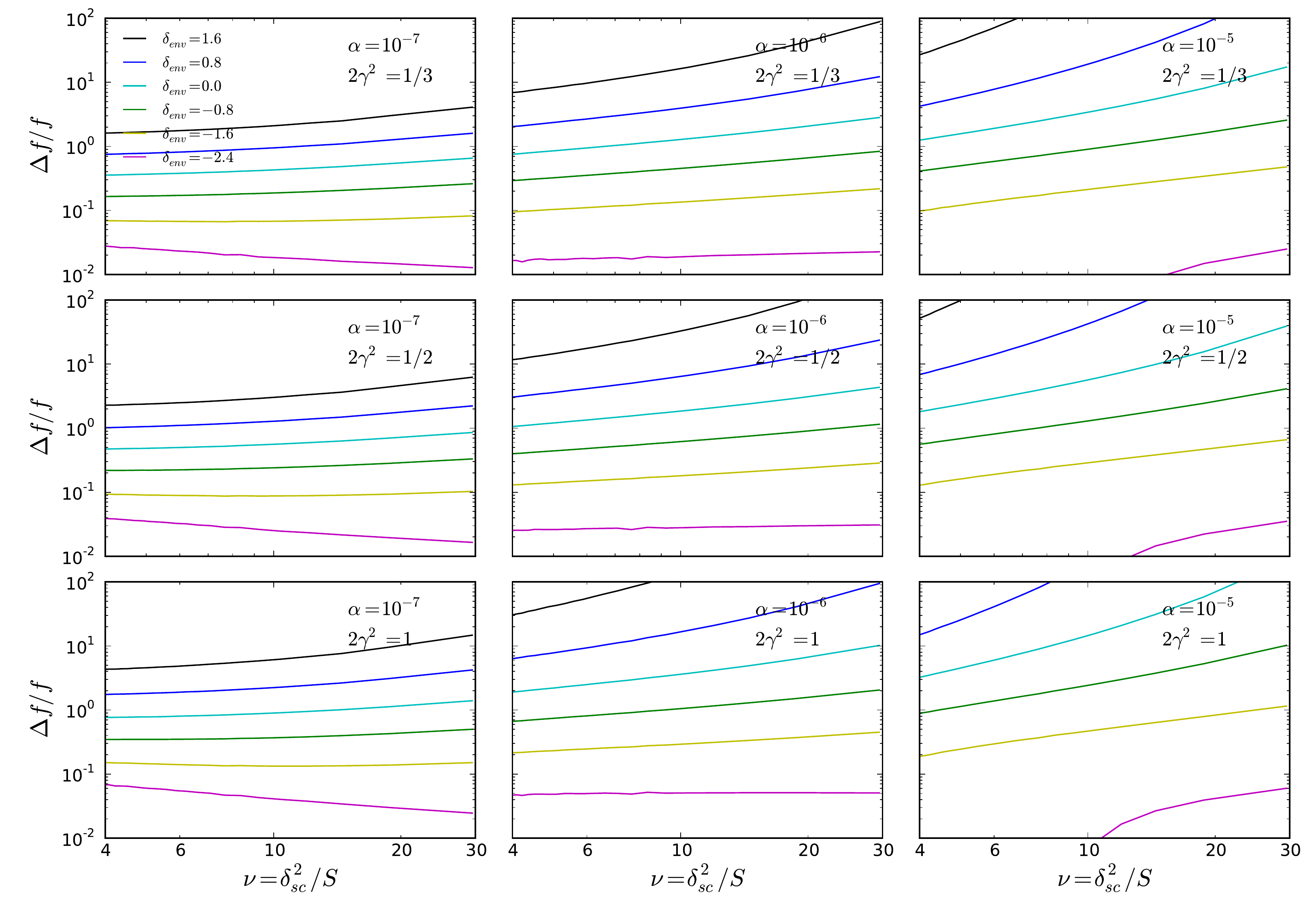}}
\caption{Fractional differences of the conditional first-crossing distribution for various parameter values. Within each panel, $\denv$ decreases from 1.6 to -2.4, from top to bottom. Our fiducial model is shown in the top center panel.}
\label{fig:diff-vary}
\end{figure*}

Figure~\ref{fig:diff-vary} shows the results for the conditional first-crossing distributions, for various parameter values. The results for any individual panel are qualitatively very similar to those for our fiducial model, $\alpha = 10^{-6}, \gamma = 1/3$. The main exception is for the $\alpha = 10^{-7}$ theories in very underdense environments, $\denv \sim -2.4$. Here the random walk begins close to a barrier which is itself very near to the $\Lambda$CDM barrier. This situation shows that the monotonic increase of the deviation with void size is not universal.

In general, larger values of $\alpha$ allow for much greater variation in the conditional first-crossing distributions for various environments. Variations in the coupling $2\gamma^2$ cause less variation between the different environments. Finally, although it is more clearly seen after the environment averaging (Fig.~\ref{fig:dndv-vary}), larger variations of the distribution with $2\gamma^2$ occur for larger $\alpha$ values.

\end{document}